\documentclass[11pt]{article}
\usepackage{graphicx}% pour inclure les figures 
\usepackage{amsmath}
\usepackage[latin1]{inputenc}
\usepackage[T1]{fontenc} 
\usepackage[french]{babel}
\usepackage{psfrag}
\newcommand{\ww}[1]{\underline{\underline{{\bf #1}}}}

\newcommand{\be}{\begin{equation}}
\newcommand{\ee}{\end{equation}}
\newcommand{\bc}{\begin{center}}
\newcommand{\ec}{\end{center}}
\newcommand{\bea}{\begin{eqnarray}}
\newcommand{\eea}{\end{eqnarray}}
\newcommand{\ba}{\begin{aligned}}
\newcommand{\ea}{\end{aligned}}
\newcommand{\bma}{\begin{bmatrix}}
\newcommand{\ema}{\end{bmatrix}}
\newcommand{\St}{{\mathcal S}}
\newcommand{\kk}{{\ww{\mathcal K}}}

\newcommand{\fext}{{\bf F}^{\text{ext}}}
\newcommand{\fint}{{\bf F}^{\text{int}}}
\newcommand{\sti }{\ww{K}}
\newcommand{\stia }{\sti^{(1)}}
\newcommand{\stib }{\sti^{(2)}}
\newcommand{\trs}[1]{\hspace{.05em}\,^{\bf T}\hspace{-.1em}\ww{#1}}
\newcommand{\rig }{\ww{G}}
\newcommand{\rigt }{\trs{G}}
\newcommand{\norm}[1]{\vert \vert {\bf #1}\vert \vert}
\newcommand{\normm}[1]{\vert \vert #1\vert \vert}
\newcommand{\bi}{\begin{itemize}}
\newcommand{\ei}{\end{itemize}}

\newcommand{\lll}{\ww{\mathcal L}}
\newcommand{\ppp}{{\mathcal P}}
\newcommand{\qqq}{{\mathcal Q}}
\newcommand{\nij}{{\bf n}_{ij}}
\newcommand{\fij}{{\bf f}_{ij}}
\newcommand{\Tij}{{\bf T}_{ij}}

\newcommand{\disty}{\displaystyle}

\newcommand{\deprel}{\vec{\mathcal U}}
\newcommand{\dep}{{\bf U}}
\newcommand{\ui}{{\bf u}_i}
\newcommand{\uj}{{\bf u}_j}
\newcommand{\ri}{{\bf r}_i}
\newcommand{\rj}{{\bf r}_j}
\newcommand{\droti}{\vec\theta_i}
\newcommand{\drotj}{\vec\theta_j}
\newcommand{\tij}{{\bf t}_{ij}}
\newcommand{\rij}{{\bf r}_{ij}}
\newcommand{\R}{\mbox{I\hspace{-.15em}R}} %l'ensemble des reels
\renewcommand{\author}[1]{
\begin{center}
{\bf #1}
\end{center}
\par
}
\renewcommand{\title}[1]{\begin{center}
    {\large{\bf #1}}
     \end{center}
\medskip
} 
\begin{document}
\title{  
M\'ethodes quasi-statiques pour la simulation num\'erique discr\`ete des assemblages granulaires }
\author{
Jean-Noël ROUX$^{(1)}$  et Ga\"el COMBE$^{(2)}$}
\bc
(1) Laboratoire Navier,
Unité Mixte de Recherche LCPC-ENPC-CNRS\\
2, allée Kepler, Cité Descartes, 77420 Champs-sur-Marne, France \\
(2) Laboratoire 3S-R,
Unité Mixte de Recherche UJF-INPG-CNRS\\ 
Domaine universitaire 2, BP53 38041 Grenoble cedex 9, France \\
\ec
\bc
\begin{minipage}{16cm}
{\small
   \parindent=0pt
    {\bf AVANT-PROPOS : }
Le pr\'esent document constitue le chapitre 3 du trait\'e  {\bf Mod\'elisation num\'erique discr\`ete des mat\'eriaux granulaires}, ouvrage collectif, sous la direction
de Farhang Radja{\"\i} et Fr\'ed\'eric Dubois,  
publi\'e dans la collection << m\'ecanique et ing\'eni\'erie des mat\'eriaux >> aux \'editions Lavoisier en 2010. Le titre, la pagination et l'indexation des r\'ef\'erences
en sont diff\'erents. Il introduit diverses d\'efinitions et propri\'et\'es relatives \`a la m\'ecanique des assemblages granulaires de type solide, pour lesquels un r\'eseau 
de contacts est \`a m\^eme de reprendre les efforts ext\'erieurement appliqu\'es et de maintenir l\'equilibre m\'ecanique sous un chargement variable, et indique comment
les calculs peuvent alors \^etre men\'es pour d\'eterminer forces et d\'eplacements sans faire appel \`a l'inertie ou \`a des forces d\'ependant du temps. Ce texte contient
quelques r\'ef\'erences 
aux autres chapitres du m\^eme trait\'e mais peut se lire ind\'ependamment.
\par
}
\end{minipage}
\ec

\section{Introduction}
Dans l'immense majorit\'e des simulations num\'eriques de mat\'eriaux granulaires \`a l'\'echelle du discret, on d\'etermine les trajectoires
d'une collection de grains en r\'esolvant les \'equations issues du principe fondamental de la dynamique, o\`u interviennent
l'inertie et les acc\'el\'erations. C'est ainsi que fonctionnent les m\'ethodes de dynamique mol\'eculaire, 
de dynamique des contacts ou pilot\'ees par \'ev\'enements qui sont
d\'ecrites dans les chapitres suivants de ce trait\'e. Pourtant dans beaucoup de situations d'int\'er\^et pratique, 
l'usage est de d\'ecrire les mat\'eriaux granulaires \`a l'\'echelle macroscopique par la m\'ecanique des milieux 
continus solides et de traiter des probl\`emes d'\'evolution 
quasi-statique, dans lesquels l'inertie n'entre pas en compte. Le syst\`eme, sous chargement variable,
passe alors par une succession d'\'etats d'\'equilibres. Les calculs aux \'el\'ements finis dans les mod\`eles 
\'elastoplastiques de la m\'ecanique des sols, par exemple, sont de ce type. Par ailleurs, les raisonnements de changement
d'\'echelle que l'on est tent\'e de b\^atir pour passer du comportement d'un assemblage de grains avec un r\'eseau de contact donn\'e
\`a une loi constitutive de mat\'eriau continu solide se fondent aussi sur une approche quasi-statique, dans laquelle l'objectif
est de d\'eterminer vers quel  \'etat 
d'\'equilibre voisin de l'\'etat initial le syst\`eme est conduit par de petits incr\'ements de forces appliqu\'ees,
 sans r\'ef\'erence au temps physique. 

Comme nous le verrons, la possibilit\'e m\^eme d'une telle \'evolution quasi-statique pose question, et c'est pourquoi les exemples d'utilisations
de m\'ethodes quasi-statiques sont encore rares dans la litt\'erature~\cite{Gael2,RC02,CR03,KTK03,McGaHe05}. 
L'approche quasi-statique se fonde sur les \emph{matrices de raideur},
\'elastiques ou \'elastoplastiques, dont la d\'efinition et la structure sont rappel\'ees au §~\ref{sec:def}, o\`u on pr\'esente \'egalement
d'autres notions fondamentales de la m\'ecanique des r\'eseaux de contact, comme les \emph{matrices de rigidit\'e} (\`a ne pas confondre avec les
matrices de raideur). On montre ensuite (§\ref{sec:disc}) comment ces diff\'erentes matrices et leurs propri\'et\'es conditionnent les 
configurations des assemblages granulaires en  \'equilibre stable, et on rappelle des observations issues des simulations de syst\`emes
de disques ou de sph\`eres qui illustrent leur importance pratique et th\'eorique. L'application des m\'ethodes 
matricielles quasi-statiques \`a l'\'elasticit\'e des assemblages granulaires, puis \`a leurs d\'eformations in\'elastiques 
 fait l'objet des §~\ref{sec:elas} et~\ref{sec:plast}. La conclusion~\ref{sec:conc} 
\'evoque bri\`evement quelques directions nouvelles vers lesquelles on pourrait appliquer ou \'etendre la m\'ethode. 

\section{Statique et cin\'ematique des r\'eseaux de contact\label{sec:def}}
On consid\'ere ici des grains en interaction par des contacts
ponctuels ou de tr\`es faible \'etendue (on supposera  par exemple que la forme des grains est r\'eguli\`ere et strictement
convexe, en excluant les contacts par des faces ou des ar\^etes de poly\`edre). 
Les mod\`eles m\'ecaniques de mat\'eriaux granulaires
utilisent en g\'en\'eral des lois de contact, qui mettent en correspondance certains \emph{mouvements relatifs} des grains avec 
les \emph{efforts de contact}. Ces correspondances peuvent \^etre lin\'earis\'ees pour de petits incr\'ements, de m\^eme que la cin\'ematique, et on fait ainsi
appara\^{\i}tre certaines matrices, dont le r\^ole est central dans l'approche quasi-statique.
Pour simplifier, on ne donnera leur \'ecriture compl\`ete que dans le cas de grains circulaires \`a deux dimensions (2D) ou sph\'eriques \`a trois dimensions (3D).
La d\'efinition de la \emph{matrice de rigidit\'e} est ici conforme 
\`a celle de la th\'eorie de la rigidit\'e (pour les treillis ou les syst\`emes 
de tens\'egrit\'e~\cite{THDU98}), que nous \'etendons aux assemblages granulaires (ce n'est pas une matrice de raideur).
\subsection{Hypoth\`ese des petites perturbations (HPP)}
\begin{figure}[!htb]
\centering
\includegraphics[width=0.6\textwidth]{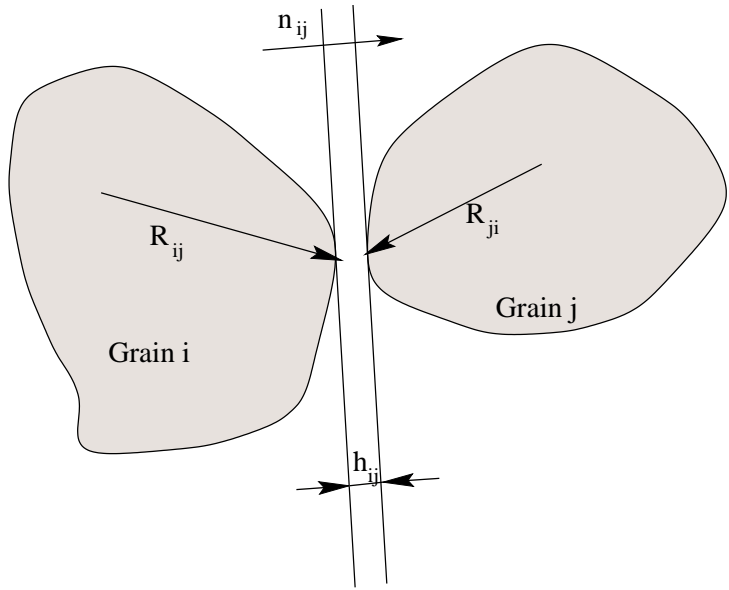}
\caption{\label{fig:contact}
Grandeurs associ\'ees \`a l'interaction entre deux grains $i$ et $j$. On d\'efinit le point de contact, ou, pour chacun des deux grains, le point de sa surface
le plus proche de son vis-\`a-vis. Ces points d\'efinissent les extr\'emit\'es des vecteurs-branches ${\bf R}_{ij}$ et ${\bf R}_{ji}$,
dont les origines respectives sont les centres (arbitrairement choisis) de $i$ et de $j$. Le vecteur unitaire ${\bf n}_{ij}$ pointe de $i$ vers $j$ et est normal
aux deux surfaces lorsqu'elles sont en contact. $h_{ij}$ d\'esigne l'interstice ou distance minimale entre les deux surfaces.
}
\end{figure}
Les approches quasi-statiques sont adapt\'ees aux calculs incr\'ementaux, dans lesquels on cherche \`a relier de petits incr\'ements de forces appliqu\'ees aux
petits d\'eplacements des grains, dont la cin\'ematique est celle d'une collection d'objets solides ind\'eformables. Sauf indication contraire on
aura recours dans la suite \`a l'\emph{hypoth\`ese des petites perturbations} (HPP), c'est-\`a-dire que l'on fera l'approximation qui consiste
\`a n\'egliger l'effet des d\'eplacements sur la g\'eom\'etrie de l'assemblage granulaire. 
On traite alors les d\'eplacements comme des vitesses, et les grandeurs g\'eom\'etriques, dont la d\'efinition est rappel\'ee sur la figure~\ref{fig:contact}, sont
gard\'ees constantes. Ainsi,
les vecteurs ${\bf R}_{ij}$, ${\bf R}_{ji}$,  ${\bf n}_{ij}$ restent fixes, tandis que $h_{ij}$ sera une fonction affine des d\'eplacements
et des (petites) rotations des grains.
Les effets de l'HPP seront discut\'es \emph{a posteriori}. Cette approximation n'est pas une vraie limitation \`a l'approche quasi-statique, qui
peut s'en dispenser. Mais on verra que les erreurs qu'elle introduit sont faibles, et nous l'adoptons aussi parce que l'expos\'e des m\'ethodes de calcul en est simplifi\'e. 
\subsection{D\'eplacements, d\'eplacements relatifs et matrice de rigidit\'e}
\begin{figure}[!htb]
\centering
\includegraphics[width=0.55\textwidth]{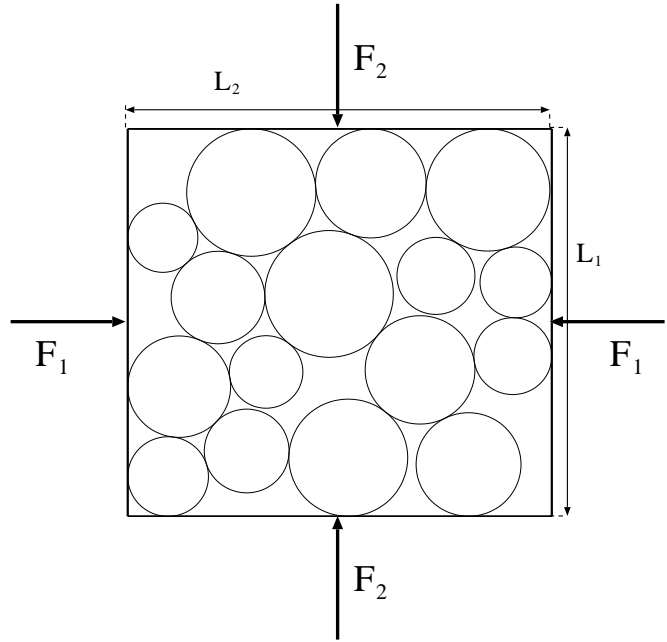}
\caption{\label{fig:biax}
Un choix de conditions aux limites en 2D adapt\'ees \`a la compression biaxiale, avec $n_g=2$ degr\'es de libert\'e associ\'es aux parois.}
\end{figure}

Nous consid\'erons, en dimension $D$ \'egale \`a deux ou trois, une collection de $N$ grains. 
Chacun d'entre eux poss\`ede $n_l = D(D+1)/2$ degr\'es de libert\'e (translation et 
rotation). Nous nous pla\c{c}ons au voisinage d'une configuration de r\'ef\'erence, 
\`a partir de laquelle les d\'eplacements sont trait\'es selon l'HPP. Les conditions aux limites
peuvent faire intervenir des objets particuliers, comme des parois, dont certains degr\'es de 
libert\'e sont fig\'es ou bien astreints \`a un mouvement ext\'erieurement impos\'e. 
Pour simplifier, on admettra que les actions ext\'erieures sur le syst\`eme consistent soit \`a interdire 
certains mouvements (cas d'une paroi fixe par exemple), soit
\`a imposer des forces (ou des moments) sur certains objets. On notera $n_g$ le nombre de degr\'es de libert\'e 
associ\'es aux conditions aux limites impos\'ees au syst\`eme. 
Le nombre total de degr\'es de libert\'e est $N_l = n_l \times N + n_g$, soit $6N+n_g$ en 3D et $3N+n_g$ en 2D.
Un exemple simple est illustr\'e par la figure~\ref{fig:biax}~:
on consid\`ere un assemblage 2D de $N$ grains, enferm\'e dans une cellule rectangulaire constitu\'ee de 4 parois, 
dont 2 (marqu\'ees 3 et 4 sur la figure) sont fixes, tandis
que celles qui leur sont oppos\'ees (respectivement~: 1 et 2) poss\`edent un seul degr\'e de libert\'e de translation dans la direction qui leur est orthogonale. Cette 
configuration permet ainsi la compression biaxiale.
Un cas int\'eressant (objet du chapitre 6 du pr\'esent ouvrage) est celui des conditions aux limites p\'eriodiques. 
On peut alors avoir des d\'eformations globales de la cellule de simulation, qui
se superposent aux mouvements des grains \`a l'int\'erieur d'une cellule p\'eriodique de forme et de taille fixes. 
Ainsi, au lieu de simuler la compression biaxiale d'un
\'echantillon 2D en le confinant entre des parois comme sur la figure~\ref{fig:biax}, on peut se donner une cellule rectangulaire p\'eriodique, 
et \'ecrire le d\'eplacement ${\bf u}_i$ de chaque grain $i$, dont la position du centre, par rapport \`a une origine quelconque 
(mais qu'il est commode de placer au centre de symétrie de la cellule), 
est rep\'er\'ee par le vecteur ${\bf r}_i$, sous la forme~:
\be
{\bf u}_i = \tilde {\bf u}_i - \ww{\epsilon}\cdot{\bf r}_i.
\label{eq:utilde}
\ee
Dans~\eqref{eq:utilde}, le tenseur de d\'eformation $\ww{\epsilon}$ (ici d\'efini selon la convention de la m\'ecanique des sols, 
c'est-\`a-dire que les raccourcissements
sont positifs) a la forme diagonale 
\be
\ww{\epsilon} = \bma 
-\frac{\delta L_1}{L_1}&0\\
0&-\frac{\delta L_2}{L_2}
\ema
\label{eq:epsbiax}
\ee
avec de petites variations $\Delta L_1$, $\Delta L_2$ des dimensions de la bo\^{\i}te, trait\'ees comme infinit\'esimales dans le cadre HPP, tandis que
$\tilde {\bf u}_i$ d\'esigne un d\'eplacement suppl\'ementaire qui satisfait aux conditions de p\'eriodicit\'e. $\Delta L_1$ et 
$\Delta L_2$ sont les $n_g=2$ degr\'es de libert\'e
associ\'es aux conditions aux limites dans ce cas.

La m\'ecanique de l'assemblage granulaire est sensible aux d\'eplacements \emph{relatifs} dans les contacts. Comme on consid\`ere de petits mouvements dans le voisinage 
imm\'ediat d'une configuration donn\'ee, on peut se donner une liste a priori de paires de grain en contact, ou \'eventuellement susceptibles d'entrer en contact. Pour 
chaque paire $i$, $j$, on choisit arbitrairement le grain origine $i$ et le grain extr\'emit\'e $j$ (comme quand on oriente les connexions sur un graphe), et le d\'eplacement
relatif $\vec{\mathcal U} _{ij}$ se d\'efinit comme la diff\'erence entre les d\'eplacements du point de contact selon le mouvement de $i$ et selon le mouvement de $j$.
Si on d\'esigne les (petites) rotations par $\droti$, $\drotj$, les d\'eplacements des centres \'etant $\ui$, $\uj$, on a alors
\be
\deprel _{ij}= \ui + \droti\times{\bf R}_{ij} -\uj - \drotj\times{\bf R}_{ji},
\label{eq:defdug}
\ee
formule g\'en\'erale faisant appara\^{\i}tre les vecteurs-branches d\'efinis sur la figure~\ref{fig:contact}, 
ce qui donne dans le cas des grains sph\'eriques ou circulaires de
rayons $R_i$ et $R_j$, en choisissant le centre conventionnel au centre g\'eom\'etrique du grain, 
\be
\deprel _{ij}= \ui -\uj  + (R_i\droti+R_j\drotj)\times\nij.
\label{eq:defdu}
\ee
Dans le cas de conditions aux limites p\'eriodiques on fera appara\^{\i}tre les d\'eplacements $\tilde\ui$ au second membre de~\eqref{eq:defdug} ou~\eqref{eq:defdu}, ainsi
qu'un terme suppl\'ementaire impliquant les d\'eformations globales au niveau de la cellule de simulation. 
Dans le petit exemple pr\'ec\'edent de la compression biaxiale 2D, on
aura ainsi, pour des grains circulaires (le vecteur rotation devenant scalaire)
\be
\deprel _{ij}= \tilde\ui -\tilde\uj  + (R_i\theta_i+R_j\theta_j)\tij +\ww{\epsilon}\cdot\rij
\label{eq:defdu2Dcircper}
\ee
o\`u $\rij$  d\'esigne la plus petite image de $\rj-\ri$ par l'une des translations qui \`a la cellule associent la famille de ses copies p\'eriodiques, 
la matrice $\ww{\epsilon}$ \'etant d\'efinie ainsi qu'en~\eqref{eq:epsbiax}, et le vecteur unitaire tangentiel $\tij$ compl\'etant $\nij$ pour former
une base directe.

Il est commode de d\'efinir un unique vecteur d\'eplacement ${\bf U}$ avec autant de coordonn\'ees que de degr\'es de libert\'e, $N_l$, en agr\'egeant les coordonn\'ees des
d\'eplacements et des vecteurs rotations de tous les grains, de $1$ \`a $N$, puis les $n_g$ degr\'es de libert\'e associ\'es 
aux conditions aux limites (d\'eplacements de parois, param\`etres de d\'eformation d'une cellule
p\'eriodique). Avec $N_c$ contacts, on d\'efinit de m\^eme un vecteur des d\'eplacements relatifs $\deprel$, 
avec $DN_c$ coordonn\'ees en dimension $D$. Il est d'usage d'utiliser
un syst\`eme de coordonn\'ees dans lequel on isole, pour chaque contact, la composante normale du d\'eplacement relatif, soit $\nij\cdot\deprel _{ij}$. Les coordonn\'ees
dans l'espace des $\deprel$ sont donc, \'etant donn\'ee a priori une liste ordonn\'ee des contacts, le d\'eplacement relatif normal, puis 
la ou les ($D-1$) coordonn\'ee(s) du d\'eplacement relatif tangentiel dans le premier contact, puis le d\'eplacement relatif normal dans le deuxi\`eme, etc., 
pour terminer par les incr\'ements des $n_g$ degr\'es de libert\'e de la cellule de simulation.

On voit alors que les relations~\eqref{eq:defdug} d\'efinissent une application lin\'eaire ${\bf U}\mapsto\deprel$,
\be
\deprel = \rig\cdot{\bf U}
\label{eq:defG}
\ee 
La matrice $\rig$ est la \emph{matrice de rigidit\'e} associ\'ee \`a l'ensemble de grains et de contacts -- la structure granulaire 
(au sens de la m\'ecanique des structures).
$\rig$ est une matrice \`a $DN_c$ lignes et $N_l$ colonnes. Le noyau de $\rig$ est l'espace des vecteurs d\'eplacements ${\bf U}$
qui ne conduisent \`a aucun d\'eplacement relatif. De tels d\'eplacements sont aussi appel\'es des mouvements de m\'ecanismes. 
La dimension de l'espace qu'ils forment, not\'ee $k$
dans la suite, est par d\'efinition le \emph{degr\'e d'hypostaticit\'e} de la structure. L'image de $\rig$ est 
le sous-espace de $\R^{DN_c}$ constitu\'e des d\'eplacements relatifs
pour lesquels il est effectivement possible de trouver des valeurs de d\'eplacements (et de rotations, etc.) qui leur correspondent. 
C'est l'espace des d\'eplacements relatifs
\emph{compatibles}, de dimension $N_l-k$.
\subsection{Efforts ext\'erieurs, forces de contact et matrice de rigidit\'e.} 
Sur chacun des grains, dont la cin\'ematique est celle d'un objet rigide, on peut appliquer une force et un moment. 
De plus, \`a chacun des $n_g$ degr\'es de libert\'e  associ\'es
aux parois ou \`a la cellule de simulation, soit $X$, on peut aussi faire correspondre une force ext\'erieure g\'en\'eralis\'ee ${\mathcal F}$, 
telle que son travail dans un
<< d\'eplacement >> $\Delta X$ soit ${\mathcal F}\Delta X$. Dans le cas du mouvement d'une paroi, comme sur la figure~\ref{fig:biax}, il s'agit d'une force
au sens le plus ordinaire. En g\'en\'eral, on peut avoir affaire \`a des forces en un sens g\'en\'eralis\'e. 
Ainsi lorsque le degr\'e de libert\'e de la cellule a le sens d'une coordonn\'ee
du tenseur des d\'eformations, $\epsilon_{\alpha\beta}$, alors la force g\'en\'eralis\'ee est 
le produit du volume par la coordonn\'ee correspondante du tenseur des contraintes,
$\sigma_{\alpha\beta}$. 
Ces efforts ext\'erieurs d\'efinissent donc un vecteur $\fext$ avec $N_l$ coordonn\'ees, le travail s'\'ecrivant $\fext\cdot{\bf U}$. 
Ils doivent \^etre \'equilibr\'es par les forces aux contacts qui d\'efinissent un vecteur ${\bf f}$ dans un espace de dimension $DN_c$. 
Par d\'efinition, la force de contact
${\bf F}_{ij}$ est la force transmise par le grain origine, $i$, au grain extr\'emit\'e, $j$, au point de contact. 
L'\'equilibre des forces g\'en\'eralis\'ees correspondant \`a tous
les degr\'es de libert\'e prend la forme d'un ensemble de relations lin\'eaires~:
\be
\fext = \ww{H}\cdot{\bf f},
\label{eq:defH}
\ee
d\'efinissant une matrice $\ww{H}$ \`a $N_l$ lignes et $DN_c$ colonnes. Le th\'eor\`eme des travaux virtuels, 
dont la v\'erification directe est facile dans les cas qui
nous int\'eressent ici, \'enonce que $\ww{H}$ n'est autre que la matrice transpos\'ee de $\rig$~:
\be
\ww{H}= \rigt.
\label{eq:virt}
\ee
La m\^eme matrice de rigidit\'e appara\^{\i}t donc dans les relations statiques et cin\'ematiques. Par exemple, dans le cas de la compression 
biaxiale d'un syst\`eme de disques avec des conditions aux limites p\'eriodiques, \'evoqu\'e plus haut, les forces g\'en\'eralis\'ees correspondant aux param\`etres de
d\'eformation $\epsilon_\alpha$, $\alpha=1$, 2 sont $A\sigma_{\alpha\alpha}$ ($\alpha=1$, 2), $A$ d\'esignant l'aire du syst\`eme dans la configuration de r\'ef\'erence.
On sait que l'on a, la somme \'etant \'etendue \`a tous les contacts (identifi\'es par la paire ordonn\'ee de disques concern\'es)~:
\be
A\sigma_{\alpha\alpha} = \sum _{i<j} r_{ij}^\alpha f_{ij}^\alpha.
\label{eq:lovebiax}
\ee
Cette relation est l'\'equation du syst\`eme lin\'eaire~\eqref{eq:defH} relative 
relative \`a force conjugu\'ee de la d\'eformation $\epsilon_\alpha$, et les coefficients de la ligne correspondante de la matrice $\ww{H}$ sont les $r_{ij}^\alpha$.
D'apr\`es~\eqref{eq:defdu2Dcircper} ce sont aussi les coefficients de la colonne de $\rig$ correspondante. 

Le noyau de $\ww{H}$ est constitu\'e des forces de contact ${\bf f}$ auto\'equilibr\'ees, c'est un sous-espace de $\R^{DN_c}$ dont la dimension $h$ est par d\'efinition le
degr\'e d'hyperstaticit\'e de la structure (ou degr\'e d'ind\'etermination des forces). 
Les forces int\'erieures \'equilibrant le chargement $\fext$ appliqu\'e, si elles existent,
forment un espace affine de dimension $h$, car \`a toute solution particuli\`ere de~\eqref{eq:defH} 
on peut ajouter une solution du syst\`eme homog\`ene associ\'e, c'est-\`a-dire
un vecteur de forces de contact auto\'equilibr\'ees. Quant \`a l'image de $\ww{H}$, c'est le sous-espace de $\R^{N_l}$ constitu\'e des chargements supportables, qu'il est
possible d'\'equilibrer avec des forces int\'erieures. 

L'image d'un op\'erateur \'etant \'egale \`a l'orthogonal du noyau de son transpos\'e, on dispose d'une caract\'erisation des d\'eplacements 
relatifs compatibles comme orthogonaux
\`a tous les syst\`emes de forces de contacts auto\'equilibr\'ees, ainsi que
d'une caract\'erisation des chargements supportables comme orthogonaux 
aux mouvements de m\'ecanisme (c'est-\`a-dire qu'ils ne travaillent pas dans ces mouvements). De plus 
on obtient une relation entre degr\'es d'hypostaticit\'e et d'hyperstaticit\'e~: 
\be
N_l + h= DN_c + k.
\label{eq:relkh}
\ee

\subsection{Lois de contact et matrices de raideur\label{sec:raid}}
Les lois de contact, dont nous supposons, conform\'ement aux mod\`eles les plus courants,
qu'elles combinent \'elasticit\'e et frottement de Coulomb, peuvent \^etre mises sous une forme incr\'ementale. Elles font alors appara\^{\i}tre des 
param\`etres de raideur. \`A chaque contact $i$,$j$ est associ\'ee une matrice de raideurs locales $\kk _{ij}$, qui relie les incr\'ements des coordonn\'ees de la
force de contact aux variations du vecteur d\'eplacement relatif~:
\be
\Delta {\bf f}_{ij}= \kk _{ij}\cdot\Delta\deprel _{ij}.
\label{eq:rloc}
\ee
Les lois de contact habituelles conduisent \`a d\'ecomposer $\fij$ en ses composantes normale et tangentielle, selon
\be
{\bf f}_{ij} = N_{ij} \nij + {\bf T}_{ij}\ \ \text{avec}\ \ {\bf T} _{ij} \perp\nij.
\label{eq:fnt}
\ee
Dans~\eqref{eq:rloc}, on choisit par cons\'equent pour les deux membres des axes de coordonn\'ees selon la direction normale pour le 
premier et selon celle de la force de contact tangentielle actuelle (avant
incr\'ementation) ${\bf T}_{ij}$ pour le deuxi\`eme, c'est-\`a-dire les 3 vecteurs de base suivants~:
\be
\nij,\ {\bf t}_{ij} = \frac{\Tij}{\normm{\Tij}},\ {\bf w}_{ij} =\nij\times{\bf t}_{ij}. 
\label{eq:base}
\ee
Dans les mod\`eles simples o\`u l'\'elasticit\'e du contact est prise lin\'eaire et unilat\'erale, 
chaque contact est dot\'e de raideurs $K_N^{ij}$, $K_T^{ij}$ ind\'ependantes des forces
qu'ils transmettent et on a pour $\kk _{ij}$ la forme diagonale simple~:
\be
\kk _{ij}^E = \bma K_N^{ij}&0&0\\0&K_T^{ij}&0\\0&0&K_T^{ij} \ema.
\label{eq:raidelas}
\ee
l'usage de la forme \'elastique~\eqref{eq:rloc}--\eqref{eq:raidelas} ne convient que si
 l'in\'egalit\'e de Coulomb est satisfaite sous forme stricte~: $\vert\vert {\bf T}_{ij}\vert\vert < \mu N_{ij}$. S'il y a \'egalit\'e dans la condition de
Coulomb, alors les raideurs locales d\'ependront de la \emph{direction} de l'incr\'ement $\Delta\deprel _{ij}$, car il faut alors discuter selon le statut 
du contact~\cite{iviso3}. On a donc~: (${\bf t}_{ij}$ est d\'efini en~\ref{eq:base})
\be
\kk _{ij} = 
\begin{cases}
\bma K_N^{ij}&0&0\\\mu K_N^{ij}&0&0\\0&0&K_T^{ij} \ema& \text{si $K_T^{ij}\Delta\deprel _{ij} \cdot {\bf t}_{ij} -
\mu K_N^{ij}\nij\cdot\Delta\deprel_{ij} > 0$ }\\
\kk _{ij}^E & \text{dans le cas contraire}\\
\end{cases}
\label{eq:raidplas}
\ee
La prise en compte de l'\'elasticit\'e de Hertz-Mindlin au contact~\cite{JO85,EB96} introduit plusieurs modifications, car
les raideurs $K_N$ et $K_T$ d\'ependent de la d\'eflexion au contact $h_{ij}$  (ou de la force normale)~\cite{JO85,EB96}, ainsi qu'en g\'en\'eral de
l'histoire du mouvement relatif tangentiel au contact. En fait la matrice des raideurs locales prend diff\'erentes
formes suivant la direction de $\Delta\deprel _{ij}$ m\^eme si le frottement n'est mobilis\'e nulle part dans la r\'egion du contact~\cite{EB96} 
(c'est-\`a-dire m\^eme si le coefficient de frottement $\mu$ est infini). Cette d\'ependance directionnelle des 
raideurs locales n'a toutefois qu'une influence tr\`es faible dans le calcul des propri\'et\'es \'elastiques macroscopiques, comme il est montr\'e dans
la r\'ef\'erence~\cite{iviso3},
o\`u on trouvera davantage de d\'etails sur la forme des matrices de raideur avec un mod\`ele de Hertz-Mindlin. Dans la suite on se limitera 
dans les exemples \`a la discussion des propri\'et\'es
\'elastiques dans le cas de grains sph\'eriques avec \'elasticit\'e de Hertz-Mindlin (§~\ref{sec:elas}) ou des propri\'et\'es \'elastoplastiques 
(§~\ref{sec:plast}) avec des disques (2D) et une \'elasticit\'e lin\'eaire unilat\'erale dans les contacts.

Quelle que soit la forme des matrices carr\'ees $\kk _{ij}$ (de taille $D\times D$) 
associ\'ees \`a chacun des contacts, on peut les rassembler dans une grande matrice carr\'ee de dimension
$DN_c\times DN_c$, la matrice des raideurs de contact $\kk$, qui est diagonale par blocs, car elle ne couple pas les contacts diff\'erents,
et qui relie des vecteurs de $\R^{DN_c}$, incr\'ements de d\'eplacements relatifs et de forces de contact~:
\be
\Delta {\bf f}= \kk\cdot\Delta\deprel.
\label{eq:rloctout}
\ee
Rappelons qu'en g\'en\'eral, la relation~\eqref{eq:rloctout} n'est lin\'eaire qu'en apparence
puisque la matrice prend des formes diff\'erentes selon la direction de $\Delta\deprel$. 
Rassemblant les \'equations~\eqref{eq:defG}, \eqref{eq:rloctout}, puis \eqref{eq:defH}
et \eqref{eq:virt}, on obtient la relation entre incr\'ements de d\'eplacements et 
de chargement ext\'erieur, qui fait appara\^{\i}tre la \emph{matrice de raideur} $\stia$~:
\be
\Delta \fext = \stia\cdot\Delta\dep, \ \ \mbox{avec }\ \ \stia = \rigt\cdot\kk\cdot\rig
\label{eq:stia}
\ee
\`A la diff\'erence de la matrice de rigidit\'e, la matrice de raideur $\stia$ est carr\'ee, avec autant de lignes et de colonnes que de degr\'es de libert\'e. 
Dans le cas o\`u chaque
bloc diagonal de $\kk$ est de la forme \'elastique~\eqref{eq:raidelas}, avec des raideurs toutes strictement positives, alors \eqref{eq:stia} montre
imm\'ediatement que $\stia$ est sym\'etrique et positive, et que son noyau co\"{\i}ncide avec celui de la matrice de rigidit\'e $\rig$.  
\subsection{Contributions g\'eom\'etriques \`a la matrice de raideur\label{sec:raidgeom}}
En ignorant les variations des directions normales et des vecteurs-branches dans les d\'eplacements, on n\'eglige, dans le cadre de l'HPP, une autre contribution
\`a la matrice de raideur -- notons-l\`a $\stib$ -- dont l'origine est la suivante.
Nous \'etudions l'effet de petits d\'eplacements au voisinage d'une configuration donn\'ee, dans laquelle l'assemblage est d\'ej\`a soumis \`a des forces ext\'erieures.
Chaque contact $i$--$j$ transmet alors une force $\fij$, qui varie lorsque les grains se d\'eplacent, d'abord en raison du mouvement relatif au point de contact, d'o\`u 
l'incr\'ement de force calcul\'e avec le comportement du contact, qui s'exprime avec la matrice de raideur $\stia$ ; mais aussi simplement parce que 
la force de contact doit suivre les grains dans leur mouvement, ce qui donne une seconde contribution \`a l'incr\'ement $\Delta\fij$ et fait appara\^{\i}tre le second terme 
$\stib$ de la  matrice de raideur. Si, par exemple, $i$ et $j$ se d\'eplacent ensemble comme un seul corps rigide, il n'y a aucun mouvement relatif au contact. 
La force $\fij$ doit alors suivre ce d\'eplacement mat\'eriel. Si le mouvement relatif de $i$ et $j$ est un roulement sans glissement, c'est-\`a-dire une rotation par
rapport \`a un axe orthogonal au vecteur normal $\nij$, ou un pivotement, c'est-\`a-dire une rotation autour de $\nij$, alors aucune variation de $\fij$ ne provient
non plus du d\'eplacement relatif, qui reste inchang\'e. La variation de la force $\fij$ dans de tels mouvements n'est curieusement presque jamais discut\'ee dans la  
litt\'erature o\`u sont pr\'esent\'ees les lois de contact et la mise en {\oe}uvre des m\'ethodes aux \'el\'ements discrets. 
Il est en g\'en\'eral suppos\'e que la force de contact
suit le mouvement de rotation du vecteur unitaire normal $\nij$ dans le roulement. Cette r\`egle est automatique
pour la composante normale $N_{ij}\nij$, dont l'intensit\'e $N_{ij}$ est g\'en\'eralement fonction de la d\'eflexion normale $h_{ij}$. On doit tenir compte du terme
$N_{ij}\Delta\nij$ d\^u \`a la rotation du vecteur unitaire normal dans le calcul de l'incr\'ement de la force de contact. Pour deux sph\`eres centr\'ees en $\ri$ et $\rj$, 
avec ${\bf r}_{ij}= \ri-\rj$ et $r_{ij}= \normm{{\bf r}_{ij}}$, on a~:
\be
\Delta \nij  = \frac{1}{r_{ij}}\left(\ww{1}-\nij \otimes
\nij \right)\cdot\left(\Delta{\bf u}_j -\Delta{\bf u}_i\right).
\label{eq:dnij}
\ee
Pour la composante tangentielle $\Tij$,
le traitement habituel des calculs 2D, dans lequel on construit un vecteur unitaire tangentiel, est similaire \`a celui de la composante normale. 
Dans les calculs 3D, il faut explicitement faire 
tourner la force tangentielle avec $\nij$, et aussi prendre en compte le pivotement. Pour ce faire, un choix naturel est d'imprimer \`a $\Tij$ la rotation moyenne des
objets $i$ et $j$ autour de $\nij$. Au total~\cite{iviso1,iviso3} cela conduit \`a incr\'ementer
la force tangentielle de $\Delta\Tij^{(2)}$ d\'efini pour deux grains sph\'eriques par la r\`egle suivante~:
\be
\Delta \Tij ^{(2)}=-\left[\Tij \cdot\left(\Delta\ui-\Delta\uj\right)\right]
\frac{\nij}{r_{ij}}
+\frac{1}{2} \left[\left(\Delta\vec\theta _i+\Delta\vec\theta _j\right)\cdot \nij\right](\nij \times \Tij).
\label{eq:dTijgeom}
\ee
Cette contribution $\Delta \Tij ^{(2)}$ est \`a ajouter au terme  $\Delta \Tij ^{(1)}$, partie tangentielle de $\Delta\fij$ calcul\'ee selon~\eqref{eq:rloctout}.
En rassemblant les contributions des composantes normales et tangentielles on trouve pour les incr\'ements de force de contact dus \`a la pr\'econtrainte 
\be
\Delta{\bf f}^{(2)} = \lll\cdot\Delta\dep,
\label{eq:dfgeom}
\ee
o\`u $\lll$ est une matrice \`a $DN_c$ lignes et $N_l$ colonnes, dont la ligne de blocs $D\times n_l$ 
relative au contact $i$-$j$ ne contient que les deux \'el\'ements non nuls $\lll _{ij,i}$ et $\lll _{ij,j}$, qui s'\'ecrivent, si on prend
la base~\eqref{eq:base} pour \'ecrire les coordonn\'ees de $\Delta\fij$ et aussi de $\Delta\dep$, et en notant
$T_{ij}=\normm{\Tij}$,
\be\ba
\lll _{ij,i} &= \bma 0&-\frac{T_{ij}}{r_{ij}}&0&0&0&0\\
0&-\frac{N_{ij}}{r_{ij}}& 0&0&0&0\\
0&0&-\frac{N_{ij}}{r_{ij}}&\frac{T_{ij}}{2}&0&0
\ema
\\
\lll _{ij,j} &= \bma 0&\frac{T_{ij}}{r_{ij}}&0&0&0&0\\
0&\frac{N_{ij}}{r_{ij}}& 0&0&0&0\\
0&0&\frac{N_{ij}}{r_{ij}}&\frac{T_{ij}}{2}&0&0
\ema
\ea
\label{eq:lij}
\ee
Pour obtenir la partie g\'eom\'etrique $\stib$ de la matrice de raideur $\sti$, 
il faut revenir \`a un syst\`eme de coordonn\'ees dans une base fixe, le m\^eme pour tous les blocs de
$\lll$, c'est-\`a-dire que chacun des deux blocs $3\times 3$ des matrices $\lll _{ij,i}$ et $\lll _{ij,j}$  
\'ecrites en~\eqref{eq:lij}, le bloc correspondant aux d\'eplacements comme le bloc correspondant aux rotations, 
est \`a multiplier \`a droite par la matrice $3\times 3$ dont les vecteurs-colonnes sont ceux de~\eqref{eq:base}.
Une fois $\lll$ ainsi transform\'ee on peut alors \'ecrire
\be
\sti = \stia + \stib,\ \text{avec $\stia$ donn\'e par~\eqref{eq:stia} et}\ \stib=\rigt\cdot\lll.
\label{eq:stibl}
\ee 
C'est en toute rigueur avec la matrice de raideur compl\`ete $\sti$ qu'il faut \'ecrire la relation entre incr\'ement de chargement et
variation de d\'eplacement plut\^ot qu'avec $\stia$ seulement
\be
\Delta\fext = \sti\cdot\Delta\dep.
\label{eq:sti}
\ee
La distance $r_{ij}$ pr\'esente dans les formules~\eqref{eq:lij} est confondue en g\'en\'eral avec la somme des rayons
$R_i+R_j$, en n\'egligeant $h_{ij}$. Comme elle provient de la variation du vecteur normal selon~\eqref{eq:dnij}, 
il s'agit des rayons de courbure des surfaces en contact et non
des vecteurs-branches qui, eux, interviennent dans la matrice de rigidit\'e selon~\eqref{eq:defdu}. Des formules g\'en\'erales
pour les coefficients des matrices $\stib$ dans le cas de grains de forme quelconque (mais r\'eguli\`ere et strictement convexe) 
sont fournies dans les r\'ef\'erences~\cite{KuCh06} et~\cite{Bagi07}. On notera que la matrice $\stib$ n'est pas sym\'etrique.
Elle le devient toutefois dans le cas particulier de grains sph\'eriques ou circulaires sans frottement, pour lesquels on peut ignorer toutes
les rotations (qui sont autant de mouvements de m\'ecanisme), car on a alors pour les blocs $D\times D$ couplant forces et d\'eplacements~:
\be
\left\{
\ba
\stib _{ij} &= \frac{N_{ij}}{r_{ij}}\left(\ww{1} - \nij\otimes\nij\right)&\text{si $j\ne i$}\hfil\\
\stib _{ii} &=-\sum_j \stib _{ij} & \text{(somme sur tous les $j$ en contact avec $i$)}
\ea
\right.
\label{eq:dksf}
\ee
D'apr\`es~\eqref{eq:lij} et \eqref{eq:stibl}, et compte tenu des coefficients de $\rig$,
les coefficients de la matrice $\stib$ couplant les forces aux d\'eplacements sont d'ordre $F/R$, o\`u $F$ est une force de contact typique et $R$ un rayon
de courbure ou une longueur de vecteur-branche, alors que les coefficients de $\stia$ correspondants sont les raideurs $K_N$, $K_T$,
multipli\'ees par des coefficients d'ordre 1 (comme les coordonn\'ees des $\nij$). 
Les coefficients qui couplent les forces aux rotations font appara\^{\i}tre un vecteur-branche suppl\'ementaire et sont de l'ordre de
$F$ (et ceux de $\stia$ d'ordre $K_N$), 
et ceux des lignes relatives aux moments comportent \'egalement un facteur $R$ suppl\'ementaire. Chacun des coefficients de $\stib$ se compare
donc \`a son analogue dans $\stia$ comme $F/R$ aux raideurs $K_N$ ou $K_T$. Or $F$ est d'ordre $K_N h$ pour une d\'eflexion de contact typique $h$, 
et on a $h\ll R$\footnote{Au chapitre 9 de ce m\^eme trait\'e on d\'efinit un param\`etre sans dimension $\kappa$ caract\'erisant le niveau de raideur de l'assemblage granulaire, 
tel que $h/R$ soit d'ordre $\kappa^{-1}$, \`a partir de la pression de confinement, du diam\`etre des grains et des raideurs de contact. 
On aura donc $K^{(2)}/K^{(1)} = O(\kappa^{-1})$ en g\'en\'eral.}, 
d'o\`u $K^{(2)}\ll K^{(1)}$ pour chacun des coefficients de la matrice couplant deux objets en contact~\cite{KuCh06,Bagi07,iviso3}. Il est donc l\'egitime de n\'egliger
la contribution g\'eom\'etrique \`a la matrice de raideur, et d'approximer~\eqref{eq:sti} par \eqref{eq:stia}, 
sauf pour les vecteurs $\dep$ tels que $\stia\cdot\dep = 0$.   
\subsection{Stabilit\'e\label{sec:stab}}
Un crit\`ere classique de stabilit\'e d'un \'etat d'\'equilibre~\cite{KuCh06} est que la forme quadratique $\Delta^2W(\Delta\dep)$, d\'efinie par
\be
\Delta^2W(\Delta\dep) = \Delta\dep\cdot\sti\cdot\Delta\dep
\label{eq:tso}
\ee
soit positive. Comme dans~\eqref{eq:tso}, $\sti\cdot\Delta\dep$ est un incr\'ement de chargement $\Delta\fext$, on peut \'ecrire la condition comme
$\Delta\fext\cdot\Delta\dep > 0$ pour tout $\Delta\dep$, c'est-\`a-dire que le << travail du second ordre >> doit \^etre positif. 
Pour comprendre l'origine de ce crit\`ere, notons que la relation~\eqref{eq:sti}, que nous avons \'ecrite en supposant l'\'equilibre de forces, 
exprime en fait que l'incr\'ement des forces \emph{int\'erieures} $\Delta\fint$ 
est \'egal \`a $-\sti\cdot\Delta\dep$. Lorsque, partant d'un état d'équilibre, une perturbation extérieure imprime aux grains des vitesses  ${\bf V}$ (rassemblées
dans un vecteur \`a  $N_l$ coordonnées, comme les petits déplacements), \`a  temps court les grains se sont déplacés de $\dep=Vt +O(t^3)$ 
(puisqu'\`a  $\dep=0$ on a une configuration d'équilibre par hypothèse) et subissent des forces intérieures 
$\Delta\fint = -\sti\cdot{\bf V}t$. La puissance  $\Delta\fint \cdot{\bf V}$ en sera strictement négative si $\Delta^2 W$ est définie positive, d'où 
une décroissance de l'énergie cinétique produite par la perturbation, qui aura tendance \`a  augmenter, en revanche, pour un vecteur ${\bf V}$, s'il existe,
tel que $\Delta^2W({\bf V}) <0$. Une telle situation est similaire à  celle d'un ressort dont la raideur serait négative, et dont la réponse tendrait à  augmenter 
l'élongation plutôt que de s'y opposer par une force de rappel.

Si $\sti$ est une matrice sym\'etrique alors on peut, pour de petits d\'eplacements $\dep$ au d\'epart d'une configuration \'equilibr\'ee, d\'ecrire les incr\'ements
de forces int\'erieures $-\sti\cdot\dep$ comme d\'erivant de l'\'energie potentielle quadratique $\Delta^2W$ (la sym\'etrie de $\sti$ assure l'\'egalit\'e des d\'eriv\'ees
secondes crois\'ees). L'\'etat d'\'equilibre est stable si l'\'energie potentielle est minimale, ce qui requiert la positivit\'e de $\Delta^2W$. 
Pour les mat\'eriaux granulaires
la sym\'etrie de $\sti$, et donc la d\'efinition d'une \'energie potentielle au voisinage d'un \'etat d'\'equilibre sous un chargement donn\'e, 
signifie qu'il existe, au moins pour
de faibles d\'eplacements, un r\'egime de comportement \'elastique. 

\section{Illustrations et discussion\label{sec:disc}}
Apr\`es le rappel de diverses notions li\'es aux r\'eseaux de contact nous 
\'evoquons dans cette section divers r\'esultats et observations num\'eriques qui 
montrent leur importance comme outils, pratiques (crit\` eres d'\'equilibre) ou th\'eoriques (formulation de probl\`emes de calcul \`a la rupture,
\'evaluation des influences de la g\'eom\'etrie du r\'eseau, de la forme des grains) d'analyse et de compr\'ehension 
de la m\'ecanique des structures de contacts intergranulaires.

\subsection{D\'efinitions, r\^oles des matrices de rigidit\'e et de raideur\label{sec:defrole}}
L'approche quasi-statique \'etant relativement peu r\'epandue, les d\'efinitions et la terminologie ne sont pas fix\'ees de fa\c{c}on unique. Ainsi les r\'ef\'erences
\cite{McGaHe05,McHe06} utilisent une matrice (dite de contact ou de configuration) not\'ee ${\bf c}$ qui est d\'efinie comme $-\rigt$ ici, et c'est à
$-\rigt$ que les auteurs de \cite{DTS05} donnent le nom de matrice de rigidit\'e. Quant à la matrice de raideur $\sti$ elle est parfois appel\'ee <<~matrice
dynamique~>>~\cite{SRSvHvS05,SvHESvS07}, comme en physique du solide. 

La matrice de rigidit\'e $\rig$ est une donn\'ee fondamentale de la structure granulaire, elle
appara\^{\i}t naturellement dans la description de la m\'ethode de dynamique des contacts, o\`u elle se combine \`a une matrice d'inertie, alors qu'elle se combine
\`a l'\'elasticit\'e du contact en dynamique mol\'eculaire (chapitre 2 de cet ouvrage) comme dans l'approche quasi-statique d\'ecrite ici. Elle ne d\'epend, par~\eqref{eq:defdug}, que
des positions des centres des grains et des points de contact. Sa transpos\'ee $\ww{H}=\rigt$ exprime l'\'equilibre des forces de
contact par~\eqref{eq:defH}. De telles forces de contact ${\bf f}$ sont dites \emph{statiquement admissibles}. 
On dit que ${\bf f}$ est plastiquement admissible si l'in\'egalit\'e
de Coulomb est satisfaite dans chaque contact. On peut alors, sans se pr\'eoccuper de la forme pr\'ecise de la loi de contact et de petites d\'eformations \'eventuelles,
s'int\'eresser \`a l'ensemble $\St$ des forces de contact \`a la fois statiquement et plastiquement admissibles. $\St$ est l'intersection de l'espace affine
de dimension $h$ (le degr\'e d'hyperstaticit\'e) des forces de contact statiquement admissibles et du c\^one des forces de contact
\emph{plastiquement admissibles}, c'est un ensemble convexe. L'approche du calcul \`a la rupture
consiste \`a d\'eclarer chargement supportable tout $\fext$ pour lequel $\St$ est non vide. Nous verrons que cette condition, certes n\'ecessaire,
ne garantit pas la stabilit\'e d'un r\'eseau de contact sous le chargement consid\'er\'e, et qu'il peut y avoir rupture avec $\St\ne\emptyset$.

Nous admettons dans ce chapitre que les efforts de contact sont des forces ponctuelles. Il arrive toutefois que, pour mod\'eliser les contacts entre grains par plusieurs
asp\'erit\'es dans le cas de surfaces rugueuses, on introduise une r\'esistance au roulement dans les contacts~\cite{IWOD98,Estrada08}. La loi de contact
correspondante, reliant un moment \`a une rotation relative peut \^etre prise analogue \`a la loi tangentielle~\cite{TOST02}, avec un coefficient de frottement 
de roulement $\mu_R$ (une longueur) qui limite la valeur du moment $\Gamma$ \`a $\mu_R N$, et une raideur en rotation. On doit alors
\'etendre la d\'efinition de ${\bf f}$, dont la dimension passe de $DN_c$ à $D(D+1)N_c/2$ pour y inclure des moments au contact (avec deux moments de roulement et un moment
de pivotement en dimension 3), tandis que $\deprel$ contiendra des rotations relatives. La r\'ef\'erence~\cite{Gilabert07} 
contient une br\`eve discussion des matrices de rigidit\'e et de raideur dans un assemblage granulaire 2D avec r\'esistance \`a la rotation.

L'\'ecriture de la matrice de raideur sous la forme $\sti = \rigt\cdot\kk\cdot\rig + \rigt\cdot\lll$ permet de d\'egager 
le r\^ole et l'influence des diff\'erentes donn\'ees g\'eom\'etriques et m\'ecaniques~: 
la structure du r\'eseau de contacts d\'etermine la matrice de rigidit\'e $\rig$, les lois de contact, une fois lin\'earis\'es pour de petits incr\'ements, 
fournissent la matrice (diagonale par blocs associ\'es \`a chacun des contacts) des raideurs de contact $\kk$, 
et la forme des grains, plus pr\'ecis\'ement leur courbure aux points de contact, n'appara\^{\i}t que dans $\lll$. Cette contribution g\'eom\'etrique \`a la matrice
de raideur intervient dans les questions de stabilit\'e. Enfin, le terme de pivotement au second membre de \eqref{eq:dTijgeom} affecte aussi la matrice $\lll$, mais
son origine n'est pas li\'ee \`a la courbure des surfaces. Il est n\'ecessaire (bien qu'en g\'en\'eral oubli\'e~!) pour assurer l'objectivit\'e~\cite{KuCh06}
du mod\`ele m\'ecanique du contact~: si les deux grains en contact sont anim\'es d'une même rotation de corps rigide autour du vecteur normal, la force tangentielle doit
subir cette rotation elle aussi. 
\subsection{Stabilit\'e et \'equilibre\label{sec:remeq}}
On est tr\`es souvent confront\'e, dans les calculs par \'el\'ements discrets \`a la question de la tol\'erance avec laquelle
des \'equations d'\'equilibre sont satisfaites. Lors de la simulation d'un essai biaxial, comme sch\'ematis\'e sur la figure~\ref{fig:biax}, le mat\'eriau est cens\'e
\^etre sollicit\'e en r\'egime quasi-statique, et \'evoluer par une suite d'\'etats d'\'equilibre. Or (comme il est indiqu\'e au chapitre 9 du pr\'esentr trait'e 
MIM), les simulations sont
toujours rapides par rapport aux essais de laboratoire. Il est classique de caract\'eriser l'\'evolution du syst\`eme dans une telle simulation au moyen de diverses
grandeurs li\'ees au r\'eseau des contacts telles que le nombre de coordination $z$ (le nombre moyen de contacts par grain), la distribution des orientations des contacts,
ou la distribution des forces. Toutes ces variables peuvent d\'ependre de l'\'ecart \`a l'\'equilibre~: un syst\`eme mal \'equilibr\'e comprendra en g\'en\'eral
moins de contacts, ceux qui portent de petites forces \`a l'\'equilibre pouvant facilement s'ouvrir sous l'effet d'une faible agitation r\'esiduelle. Lorsque l'on cherche
\`a savoir si le r\'eseau des contacts est correctement d\'etermin\'e, il est utile de tester la stabilit\'e de l'\'equilibre au moyen du crit\`ere de positivit\'e
de la forme quadratique~\eqref{eq:tso}. Cette op\'eration se trouve facilit\'ee si la contribution g\'eom\'etrique $\stib$ est n\'egligeable (absence de m\'ecanisme)
et si le frottement n'est pas mobilis\'e. Or c'est effectivement ce qui arrive, comme on l'observe dans la pratique des simulations, du moins avec 
des disques ou des sph\`eres~: on constate que le nombre de contacts glissants tend à diminuer et à s'annuler \`a mesure que les \'ecarts \`a l'\'equilibre
se r\'eduisent. Comme le syst\`eme approche de son \'equilibre final, il est anim\'e de diff\'erents modes de vibrations, dans lesquelles les forces de contact ${\bf f}$
oscillent de fa\c{c}on assez erratique, et, g\'en\'eriquement, l'arr\^et de ces oscillations en un point de l'ensemble $\St$ des forces statiquement et plastiquement
admissibles ne se produit pas sur sa fronti\`ere (l\`a o\`u l'in\'egalit\'e de Coulomb est satur\'ee dans au moins un contact).  
La positivit\'e de la matrice de raideur, alors en bonne approximation sym\'etrique, est assur\'ee si les mouvements de m\'ecanisme n'affectent pas
la structure qui porte les forces. Pour choisir la tol\'erance sur l'\'equilibre on peut prendre une valeur de force telle que l'on obtienne
avec des \'ecarts plus faibles des matrices de raideur d\'efinies positives. On a ainsi observ\'e, pour des assemblages de billes~\cite{iviso1} qu'une tol\'erance
de l'ordre de $10^{-4}F_N$,  o\`u  $F_N$ est la force de contact normale moyenne, est en g\'en\'eral assez faible pour cela. Si on impose que le r\'esultante des forces 
sur chaque grain soit inf\'erieure \`a $10^{-4}F_N$, que le moment r\'esultant soit inf\'erieur \`a $10^{-4}F_Nd$ o\`u $d$ est le diam\`etre du grain, tandis que les
contraintes ext\'erieurement impos\'ees sont \'equilibr\'ees par les forces int\'erieures avec une erreur relative inf\'erieure \`a $10^{-4}$, alors on v\'erifie que
le r\'eseau des contacts d\'efinit une matrice de raideur qui assure la stabilit\'e su syst\`eme. 
\subsection{Hyperstaticit\'e, hypostaticit\'e, isostaticit\'e et raret\'e des contacts\label{sec:rare}}
Les r\'eseaux de contact des assemblages granulaires \`a l'\'equilibre, dans le cas de grains circulaires (2D) ou sph\'eriques, ne poss\`edent en g\'en\'eral 
que des mouvements de m\'ecanisme simples qu'il est facile d'\'eliminer lors de la construction des matrices de rigidit\'e, quitte \`a diminuer la dimension 
$N_l$ de l'espace des vitesses ou des petits d\'eplacements. Ainsi, dans le cas d'une cellule p\'eriodique, les translations d'ensemble (mais pas les
rotations) sont des m\'ecanismes triviaux, au nombre de $D$. En g\'en\'eral notons $k_0\le D(D+1)/2$ 
le nombre de m\'ecanismes qui sont des mouvements d'ensemble de corps rigide.  Une structure qui ne poss\`ede pas d'autre mouvement de m\'ecanisme est dite rigide (ou, plus 
pr\'ecis\'ement, rigide au premier ordre~\footnote{En g\'en\'eral la propri\'et\'e de rigidit\'e au premier ordre est plus forte que la seule rigidit\'e,
qui est l'impossibilit\'e de d\'eformer la structure sans qu'il y ait d\'eplacement relatif dans un contact~\cite{THDU98}.}).

D'autres m\'ecanismes \'evidents sont les mouvements des grains flottants, c'est-à-dire qui
ne transmettent aucune force. Dans les assemblages de disques ou de sph\`eres faiblement polydispers\'es, la proportion $x_0$ de grains flottants, selon la mani\`ere
de pr\'eparer l'\'etat d'\'equilibre peut varier de proche de z\'ero \`a plus de 15\%~\cite{SRSvHvS05,iviso1}. Avec une grande \'etendue granulom\'etrique~\cite{thVoivret},
le nombre de flottants parmi les grains de petite taille peut \^etre beaucoup plus \'elev\'e. Tous les degr\'es de libert\'e des grains flottants sont des m\'ecanismes,
d'o\`u $k\ge n_lx_0N$. 

Beaucoup d'\'etudes ont port\'e sur le cas des disques ou des sph\`eres sans frottement \cite{Gael2,RC02,OSLN03,DTS05,iviso1}. Toutes les rotations
sont alors des m\'ecanismes, et on peut en fait les retirer de la liste des degr\'es de libert\'e car elles ne changent \'evidemment pas la g\'eom\'etrie
du syst\`eme et ne donnent lieu \`a aucun d\'eplacement relatif normal dans les contacts. On peut retrancher $2DN_c$ \`a la dimension des espaces de forces
de contact ou de d\'eplacements relatifs (en supprimant les composantes tangentielles), retrancher $N$ \`a $k$, et remplacer $N_l$ par 
$N_l^{(0)}=N_l - D(D-1)N/2$ (en supprimant les rotations), de sorte que \eqref{eq:relkh} devient
\be
N_l^{(0)} + h= N_c + k.
\label{eq:relkh00}
\ee
Pour des objets non frottants de forme g\'en\'erale, les rotations ne sont plus a priori des mouvements de m\'ecanisme, et on aura simplement
\be
N_l + h= N_c + k,
\label{eq:relkh0}
\ee
tandis que pour des objets de r\'evolution (3D) on a une rotation libre par grain et~\eqref{eq:relkh00} s'applique avec $N_l^{(0)}=N_l - N$.

La forme \eqref{eq:dksf} de $\stib$ dans ce cas est une matrice sym\'etrique n\'egative, ce qui montre que tous les m\'ecanismes dans lesquels il y a un
d\'eplacement relatif normal non nul dans au moins un contact conduisent \`a des instabilit\'es. Si on a obtenu un \'etat d'\'equilibre stable, alors on doit avoir 
$k=k_0 + DN_0$ dans \eqref{eq:relkh0}, $N_0$ \'etant le nombre de grains flottants. Dans un grand syst\`eme, ceci entra\^{\i}ne une in\'egalit\'e pour le
nombre de coordination $z^*$ de l'ensemble des $N^*=N-N_0$ grains portant des forces \`a l'\'equilibre ($z^*=z/(1-x_0)$). En prenant $N_C=2z^*N^*$, \eqref{eq:relkh0}, o\`u
$k=k_0$  est n\'egligeable et $h\ge0$ donne $z^* \ge 6$ pour des sph\`eres (3D) et $z^* \ge 4$ pour des disques (2D).
En revanche, des grains de forme ellipso\"{\i}dale peuvent former sans frottement des assemblages stables avec des mouvements de m\'ecanismes possibles non 
triviaux~\cite{Donev-ellipse}, la matrice $\stib$ pouvant, selon la courbure des surfaces aux points de contact~\cite{KuCh06}, 
stabiliser ces mouvements ce qui autorise, selon~\eqref{eq:relkh}, des valeurs $z^*<12$. 

\begin{figure}[!htb]
\centering
\includegraphics[width=0.65\textwidth]{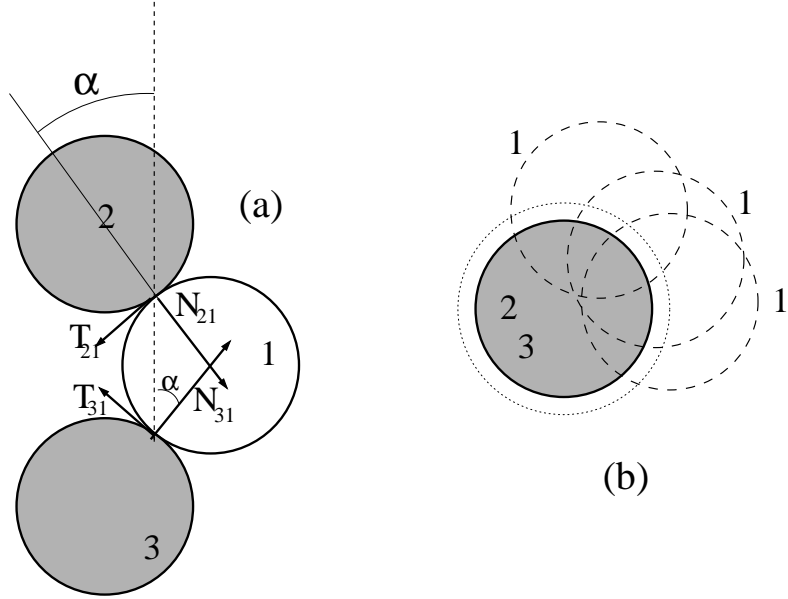}
\caption{\label{fig:dival} Mouvement de m\'ecanisme d'une sph\`ere divalente (num\'ero 1), 
la droite joignant ses deux points de contact avec les sph\`eres 2 et 3, immobiles,
est l'axe instantan\{e de rotation et porte les 2 forces de contact. 
(a) \'Equilibre des forces sur la sph\`ere 1, dans le plan d\'efini par les trois centres. 
L'\'equilibre est possible si $\alpha$ est inf\'erieur \`a l'angle de frottement
de contact. (b) Mouvement,
vu de dessus (2 et 3 sont en position \'eclips\'ee). Le centre de la sph\`ere mobile 1 
d\'ecrit le cercle en pointill\'es fins autour de l'axe joignant les centres de 2 et 3.
}
\end{figure}

Avec des grains frottants, l'unique m\'ecanisme non trivial observ\'e pour des billes sph\'eriques est celui de la figure~\ref{fig:dival}, qui met en mouvement 
les particules \`a deux contacts, le reste de l'assemblage restant fixe~\cite{iviso1}. 
En chacun des deux contacts de la sph\`ere mobile, il s'agit d'une combinaison de roulement et de
pivotement, dans laquelle les forces de contact restent constantes, alors que les points de contact changent et
d\'ecrivent sur la surface des grains une trajectoire circulaire.
On v\'erifie que dans ce mouvement la contribution des contacts de la bille mobile au travail du second ordre est nul, le vecteur vitesse ${\bf V}$ 
est tel que $\stia\cdot{\bf V}=\stib\cdot{\bf V}=0$. 
La stabilit\'e ou l'instabilit\'e ne se manifeste donc qu'au travers des effets d'une variation de moment subi de la part de la bille marqu\'ee 1 sur la figure par les billes
marqu\'ees 2 et 3. Les simulations donnent une population faible (2 ou 3\%) mais non n\'egligeable de telles particules divalentes dans les assemblages faiblement
coordonn\'es~\cite{iviso1}, les forces qu'elles transmettent peuvent \^etre importantes, et la proportion de m\'ecanismes instables tend à dispara\^{\i}tre 
quand la pression de confinement augmente. Si on trouve des grains de forme non sph\'erique avec deux contacts dans un assemblage 3D, on peut \'egalement leur associer un
mouvement de m\'ecanisme similaire, mais l'occurrence de telles configurations et leur stabilit\'e ne semble pas encore avoir \'et\'e r\'epertori\'ees dans la litt\'erature.

En admettant l'absence de m\'ecanismes non triviaux autres qu'associ\'es aux grains divalents, en proportion
$x_2^*$ parmi les $N^*=N(1-x_0)$ grains qui transmettent des forces, la relation~\eqref{eq:relkh}
permet de minorer la coordinence $z^*=2N_c/N^*$ du r\'eseau des contacts actifs
\be
\ba
z^* &\ge 4 - \frac{2}{3}x_2^*&\text{(grains sph\'eriques 3D, $\mu=0$)}\\
z^* &\ge 3&\text{(grains circulaires 2D, $\mu=0$)}\\
\ea
\label{eq:minz}
\ee

Pour le degr\'e d'hyperstaticit\'e, on dispose \'egalement de r\'esultats pour les grains non frottants~\cite{MO98a,TW99,JNR2000,MO01}. 
En effet, pour des configurations g\'en\'eriquement
d\'esordonn\'ees (en pratique, pour tous les assemblages sauf les r\'eseaux parfaitement ordonn\'es), le degr\'e d'hyperstaticit\'e des assemblages de grains non frottants
est nul dans la limite rigide des faibles contraintes de confinement, ou des grandes raideurs de contact\footnote{C'est la limite o\`u 
$\kappa\to+\infty$, $\kappa$ \'etant le
param\`etre de raideur d\'efini au chapitre 9.}. Une situation analogue famili\`ere est celle de la table \`a quatre pieds qui en g\'en\'eral est bancale si ses
contacts avec le sol sont rigides, parce
que les incertitudes g\'eom\'etriques sur la forme des pieds comme sur les irr\'egularit\'es du sol interdisent la configuration hyperstatique \`a quatre contacts.
Cette propri\'et\'e d'absence d'hyperstaticit\'e est en fait de nature g\'eom\'etrique -- le r\'eseau des contacts ne peut pas supporter un syst\`eme de forces
normales auto-\'equilibr\'ees -- et sa validit\'e est ind\'ependante de la valeur effective du coefficient de frottement intergranulaire. Une cons\'equence imm\'ediate en est
une majoration du nombre de coordination $z^*$ de l'assemblage priv\'e de ses grains flottants dans la limite rigide. 
\`A partir des relations~\eqref{eq:relkh00} ou ~\eqref{eq:relkh0}, 
on obtient en effet~:
\be
\ba
z^*&\le 2D &\ \text{(disques ou sph\`eres, $D=2$ ou 3)}\\
z^*&\le D(D+1)/2 &\ \text{(grains quelconques, $D=2$ ou 3)}\\
z^*&\le 10  &\ \text{(grains axisym\'etriques, $D=3$)}
\ea
\label{eq:majz}
\ee
\`A la diff\'erence de~\eqref{eq:minz}, ces in\'egalit\'es restent vraies quel que soit le coefficient de frottement $\mu$, mais seulement dans la limite des contacts rigides
et ind\'eformables.

La structure des grains en contact portant des forces est \emph{isostatique} quand elle est d\'epourvue d'hyperstaticit\'e ($h=0$) et d'hypostaticit\'e, sauf les
\'eventuels mouvements d'ensemble de corps rigide ($k=k_0$). La matrice $\rig$, si on restreint 
l'espace des vitesses ou des petits d\'eplacements en excluant de tels mouvements,
est carr\'ee et inversible. C'est effectivement le cas pour des billes ou des disques rigides non frottants (et non coh\'esifs), 
en pr\'esence de d\'esordre g\'en\'erique. On a alors
$z^*=6$ pour les billes, $z^*=4$ pour les disques, d'apr\`es \eqref{eq:minz} et \eqref{eq:majz}. 
Ces propri\'et\'es sont effectivement observ\'ees dans la limite rigide~\cite{CR2000,Gael2,OSLN03,DTS05,iviso1}. L'isostaticit\'e
est propre aux grains circulaires ou sph\'eriques, des objets de forme diff\'erente pouvant s'assembler dans des configurations avec des m\'ecanismes 
stables~\cite{Donev-ellipse}. L'isostaticit\'e a \'et\'e exploit\'ee pour mettre au point des m\'ethodes de calcul quasi-statique d'assemblages de disques rigides se
r\'earrangeant sous chargement variable sans aucun autre param\`etre que g\'eom\'etrique~\cite{CR2000}.
L'absence d'hyperstaticit\'e s'applique plus g\'en\'eralement dans la limite rigide. 
Elle a des cons\'equences remarquables (les forces ne d\'ependent pas de la loi
de contact) mais ne vaut pas pour les grains frottants, même dans la limite rigide~\cite{iviso1}, sauf \'eventuellement pour certains
proc\'ed\'es d'assemblage dans la limite $\mu\to\infty$ (qui est une curiosit\'e th\'eorique).
Il a parfois \'et\'e sugg\'er\'e qu'on pourrait avoir une sorte
d'isostaticit\'e << g\'en\'eralis\'ee >> pour des grains frottants \`a l'approche d'une rupture d'un assemblage, 
en donnant une coordonn\'ee de moins au vecteur des forces de contact
l\`a o\`u il y a glissement et donc \'egalit\'e dans la condition de Coulomb. Cette pr\'ediction est toutefois contredite par
les observations num\'eriques~\cite{Gael2}. De plus, cette notion d'ind\'etermination des forces prenant en compte le statut des contacts est d'un usage plus d\'elicat,
car elle d\'epend des forces appliqu\'ees et ne correspond \`a aucune propri\'et\'e duale pour les d\'eplacements. 

On notera que ces questions d'hyperstaticit\'e ou d'hypostaticit\'e ne prennent aucunement en compte les conditions in\'egalit\'es qui portent sur les forces. Cependant,
la majoration \eqref{eq:majz} du nombre de contacts dans un assemblage granulaire d\'esordonn\'e entra\^{\i}ne une certaine limitation du degr\'e d'hyperstaticit\'e, 
qui est li\'e \`a certaines propri\'et\'es assez g\'en\'erales des mat\'eriaux granulaires (comme la distribution des forces), et tend \`a restreindre l'influence de
la loi de contact et \`a renforcer celle de la g\'eom\'etrie. Ainsi, en g\'en\'eral, si l'on conna\^{\i}t avec pr\'ecision les positions (et \'eventuellement les
orientations) des grains, on peut calculer, avec les lois les plus habituelles, les forces de contact normales, qui sont \'elastiques et li\'ees \`a la d\'eflexion des
contacts (qui appra\^{\i}t dans une simulation num\'erique comme une << interp\'en\'etration >>). En revanche la force tangentielle d\'epend de l'histoire
des sollicitations du contact et n'est pas directement d\'etermin\'ee par les positions actuelles. C'est pourquoi, dans la pratique des calculs par \'ele\'ements discrets,
on ne sauvegarde pas seulement les positions mais aussi les forces \`a la fin du calcul. Cependant, on peut montrer que dans un \'etat d'\'equilibre les forces tangentielles
sont en g\'en\'eral uniquement d\'etermin\'ees pour les disques ou les sph\`eres satisfaisant aux in\'egalit\'es~\eqref{eq:majz}. 
Dans le cas de sph\`eres, les \'equations d'\'equilibre sont
au nombre de $N_l=6N^*$ et contiennent $z^*N^*$ coordonn\'ees inconnues de forces tangentielles (c'est-\`a-dire 2 pour chacun des $z^*N^*/2$ contacts), et celles-ci sont donc,
g\'en\'eralement, d\'etermin\'ees. Une cons\'equence en est que dans bien des \'evolutions
quasi-statiques, l'oubli du dernier terme de~\eqref{eq:dTijgeom}, qui viole le principe d'objectivit\'e~\cite{KuCh06} peut s'av\'erer inoffensif, la stabilit\'e impliquant
finalement un retour \`a l'\'equilibre avec les m\^emes valeurs des forces tangentielles.
\section{Applications \`a l'\'elasticit\'e\label{sec:elas}}
Nous rappelons ici quelques r\'esultats relatifs aux propri\'et\'es \'elastiques \'evalu\'ees num\'eriquement pour des assemblages de billes pour lesquels l'\'elasticit\'e des
contacts ob\'eit aux lois de Hertz-Mindlin-Deresiewicz~\cite{JO85}. Des comparaisons avec les r\'esultats exp\'erimentaux sont possibles, pour les valeurs des modules ainsi
que pour l'extension du domaine de comportement (approximativement) \'elastique. Elles donnent de bons r\'esultats pourvu que les \'etats internes
du mat\'eriau dans l'exp\'erience et dans la simulation soient proches, en particulier les nombres de coordination et les tenseurs de texture (orientation
des contacts). Nous renvoyons \`a~\cite{Emam06,iviso3} pour les r\'ef\'erences aux travaux exp\'erimentaux
sur les sables ou sur les billes et pour le d\'etail des confrontations
entre simulations et exp\'eriences. L'\'elasticit\'e relie de petits incr\'ements de contrainte \`a de tr\`es faibles d\'eformations 
au voisinage imm\'ediat d'un \'etat de r\'ef\'erence, et pourrait de ce fait \^etre consid\'er\'ee comme peu pertinente pour le comportement m\'ecanique des mat\'eriaux
granulaires. Toutefois, les modules \'elastiques sont accessibles \`a l'exp\'erience et fournissent indirectement des mesures non destructives de caract\'eristiques importantes
des assemblages granulaires comme le nombre de coordination, qui, le plus souvent, 
\'echappent aux techniques d'observation directe~\cite{iviso3}. La simulation de la r\'eponse
\`a de petites variations de chargement est en outre justifi\'ee par l'\'etude de crit\`eres de localisation de la d\'eformation~\cite{VaSu95}, qui demandent une connaissance
pr\'ecise de la loi constitutive sous forme incr\'ementale. Dans cette partie notre propos est 
d'illustrer les apports de la m\'ethode quasi-statique. Nous ne traitons pas des valeurs des modules, de leur d\'ependance par rapport aux contraintes et
\`a la structure de l'assemblage~\cite{Emam06,iviso3}, ni de leur pr\'ediction, 
ou des propri\'et\'es quelque peu anormales des r\'eseaux faiblement hyperstatiques~\cite{Wyart-th2,SvHESvS07,iviso3}.
\subsection{Calcul des modules \'elastiques}
De petits d\'eplacements donneront des forces \'elastiques, comme on l'a vu au §~\ref{sec:stab}, si on peut \'ecrire une matrice de raideur sym\'etrique d\'efinie
positive. La cause majeure d'asym\'etrie dans la matrice de raideur est la mobilisation du frottement, Eq.~\eqref{eq:raidplas}. Lorsque l'on a affaire \`a un assemblage
bien \'equilibr\'e  on a en pratique $\normm{\Tij}< \mu N_{ij}$ dans tous les contacts $i$-$j$ (§~\ref{sec:remeq}), de sorte que l'on peut garder la forme sym\'etrique et
d\'efinie positive de chacun des blocs diagonaux de $\kk$. 
L'\'elasticit\'e suppose aussi que l'on puisse n\'egliger la contribution g\'eom\'etrique non sym\'etrique \'evalu\'ee (pour des sph\`eres) au §~\ref{sec:raidgeom}, 
ce qui est possible s'il n'y a pas de m\'ecanisme. Lorsqu'il s'agit de calculer une r\'eponse \'elastique, on ignore bien s\^ur les grains flottants.
Restent les m\'ecanismes associ\'es aux grains divalents (figure~\ref{fig:dival}), qui appartiennent au noyau de $\stia$ et \`a celui de $\stib$. \`A moins d'exercer
directement des efforts sur la particule divalente (marqu\'e 1 sur la figure~\ref{fig:dival}), un incr\'ement de chargement, en particulier une variation de contrainte
globale, ne travaille pas dans ces m\'ecanismes, que l'on peut \'eliminer soit en r\'eduisant le nombre de degr\'es de libert\'e, soit en attribuant une raideur finie 
au mouvement libre, comme si, par exemple, le centre de la particule mobile \'etait maintenue dans le plan de la figure~\ref{fig:dival}(a) par un ressort.
Enfin, une derni\`ere cause d'asym\'etrie de la matrice de raideur r\'eside dans la sensibilit\'e des champs de contraintes et de d\'eformation dans la r\'egion du contact
entre deux grains au trajet de chargement~\cite{EB96}. Il a \'et\'e v\'erifi\'e que ces subtilit\'es de la loi de contact n'ont qu'un tr\`es faible impact sur le calcul
des modules~\cite{iviso3}. On peut donc \`a partir d'une configuration bien \'equilibr\'ee construire la matrice de raideur \'elastique sous la forme $\stia$ avec la forme
\'elastique~\eqref{eq:raidelas} de la matrice des raideurs locales $\kk$. Apr\`es \'elimination \'eventuelle des m\'ecanismes associ\'es aux grains divalents, on résout le
syst\`eme lin\'eaire 
\be
\sti\cdot\dep = \Delta\fext,
\label{eq:systelas}
\ee
dont l'inconnue est le vecteur d\'eplacement $\dep$, et l'incr\'ement $\Delta\fext$ au second membre a des coordonn\'ees nulles sauf celles qui expriment un incr\'ement de
contrainte (ou de force sur une paroi). Alternativement, on peut imposer des d\'eformations de l'\'echantillon en manipulant les conditions aux limites, ce qui revient
\`a imposer certaines coordonn\'ees de $\dep$. En partitionnant les degr\'es de libert\'e entre ceux qui sont impos\'es, coordonn\'ees d'un vecteur ${\bf u}_g$ de dimension
$n_g \ll N^*$ et ceux qui sont laiss\'es libres, coordonn\'ees de $\tilde\dep$ de dimension $N_l^*-n_g$, 
ainsi que la matrice $\sti$ en blocs correspondants, \eqref{eq:systelas}
prend la forme
\be
\bma
\tilde\sti&\ww{L}\\
\trs{L}&\ww{k}_g
\ema\cdot
\bma
\tilde\dep\\
{\bf u}_g
\ema
=\bma
\Delta\tilde\fext\\
\Delta {\bf f}_g
\ema
.
\label{eq:kblocs}
\ee
On trouve alors le syst\`eme d'\'equations \`a r\'esoudre pour $\tilde\dep$ en prenant $\Delta\tilde\fext=0$ dans \eqref{eq:kblocs}, o\`u ${\bf u}_g $ est connu~:
\be
\tilde\sti\cdot\tilde\dep=-\ww{L}\cdot{\bf u}_g.
\label{eq:systelasred}
\ee
Que l'on utilise \eqref{eq:systelas} ou bien \eqref{eq:systelasred},
on doit r\'esoudre un syst\`eme lin\'eaire avec une matrice sym\'etrique positive d\'efinie et creuse. Pour ce faire, on dispose de diverses m\'ethodes, dont celle
du gradient conjugu\'e~\cite{Golub93} (\'eventuellement pr\'econditionn\'e), 
peu co\^uteuse en m\'emoire mais pouvant donner de longs calculs si la matrice de raideur est
mal conditionn\'ee, ou bien la factorisation de Cholesky, plus co\^uteuse pour un seul syst\`eme lin\'eaire, 
et pour laquelle il faut tenter de minimiser le stockage m\'emoire
n\'ecessaire en r\'eordonnant au besoin les inconnues, mais qui finit par \^etre avantageuse si on doit r\'esoudre des syst\`emes avec la m\^eme matrice mais 
pour de nombreuses valeurs diff\'erentes du second membre. 
Les matrices de raideur des assemblages granulaires ayant une structure tr\`es similaire \`a celles que l'on rencontre dans les probl\`emes d'\'elasticit\'e 
discr\'etis\'es par \'elements finis, on trouvera des indications utiles sur la r\'esolution num\'erique dans la litt\'erature plus vaste qui leur est consacr\'ee.

Bien entendu, il est possible de calculer des modules \'elastiques sans recourir \`a 
la matrice de raideur, en simulant par les m\'ethodes dynamiques habituelles la
r\'eponse \`a de petits incr\'ements de contrainte appliqu\'ee. Il faut ensuite s'assurer qu'il y a bien une r\'eponse lin\'eaire dans un certain 
intervalle de sollicitations, car des incr\'ements de chargement trop faibles donnent des r\'esultats affect\'es par les petits \'ecarts \`a l'\'equilibre,
et des incr\'ements trop grands sortent du domaine \'elastique. 
Un tel calcul renseigne aussi sur le domaine \'elastique (ou approximativement \'elastique). Le recours
\`a la matrice de raideur~\cite{SRSvHvS05,iviso3,SvHESvS07} donne cependant un acc\`es plus direct et plus rapide \`a l'ensemble des modules. 
\subsection{Le domaine \'elastique}  
\begin{figure}[!htb]
\centering
\includegraphics[angle=270,width=.48\textwidth]{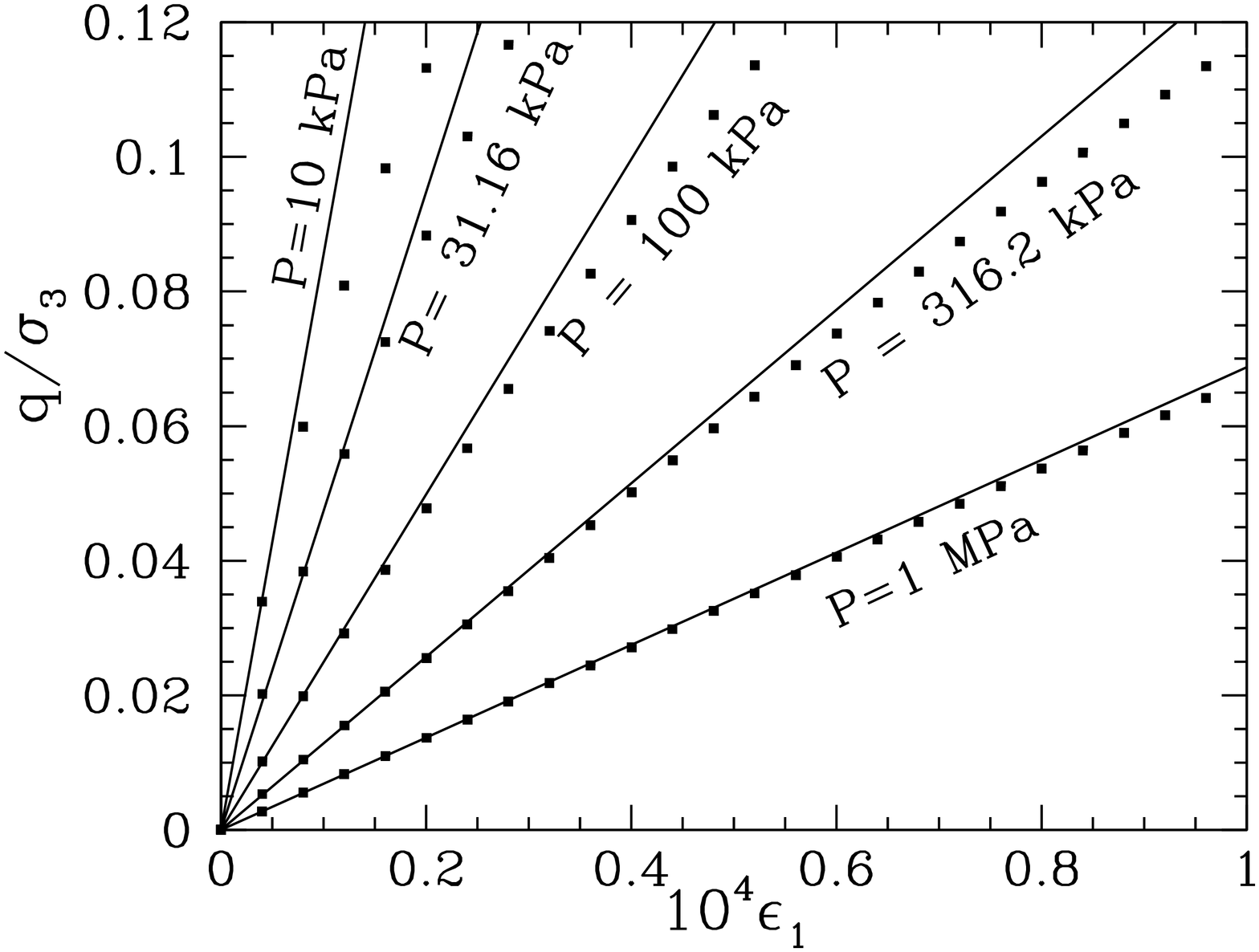}
\hfil
\includegraphics[angle=270,width=.48\textwidth]{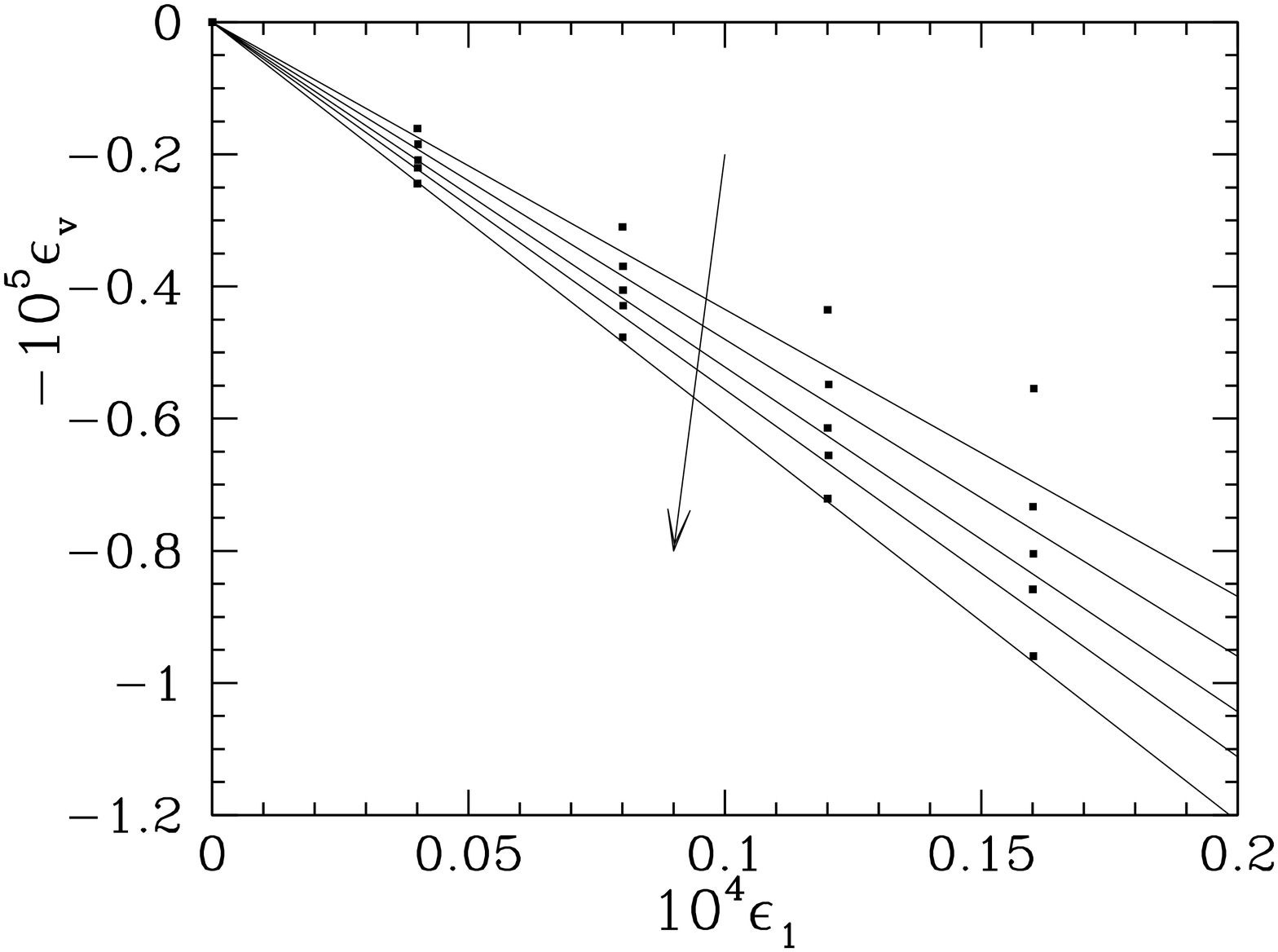}
\caption{\label{fig:prq}
D\'eviateur des contraintes (\`a gauche), normalis\'e par la contrainte lat\'erale $\sigma_3=P$, et d\'eformation volumique (\`a droite)
fonctions de la d\'eformation axiale dans la compression triaxiale simul\'ee d'un assemblage de billes de verre pour 5
valeurs de $P$. Les points donnent les r\'esultats du calcul par dynamique
mol\'eculaire, et les droites en lignes continues ont pour pentes $E^*$ et $(1-2\nu^*)$,  $E^*$ et $\nu^*$ \'etant le module d'Young et le coefficient de Poisson 
du mat\'eriau granulaire dans l'\'etat initial isotrope. La fl\`eche indique pour les courbes de $\epsilon_v$ le sens de $P$ croissant.
}
\end{figure}

\begin{figure}[!htb]
\centering
\includegraphics[width=.48\textwidth]{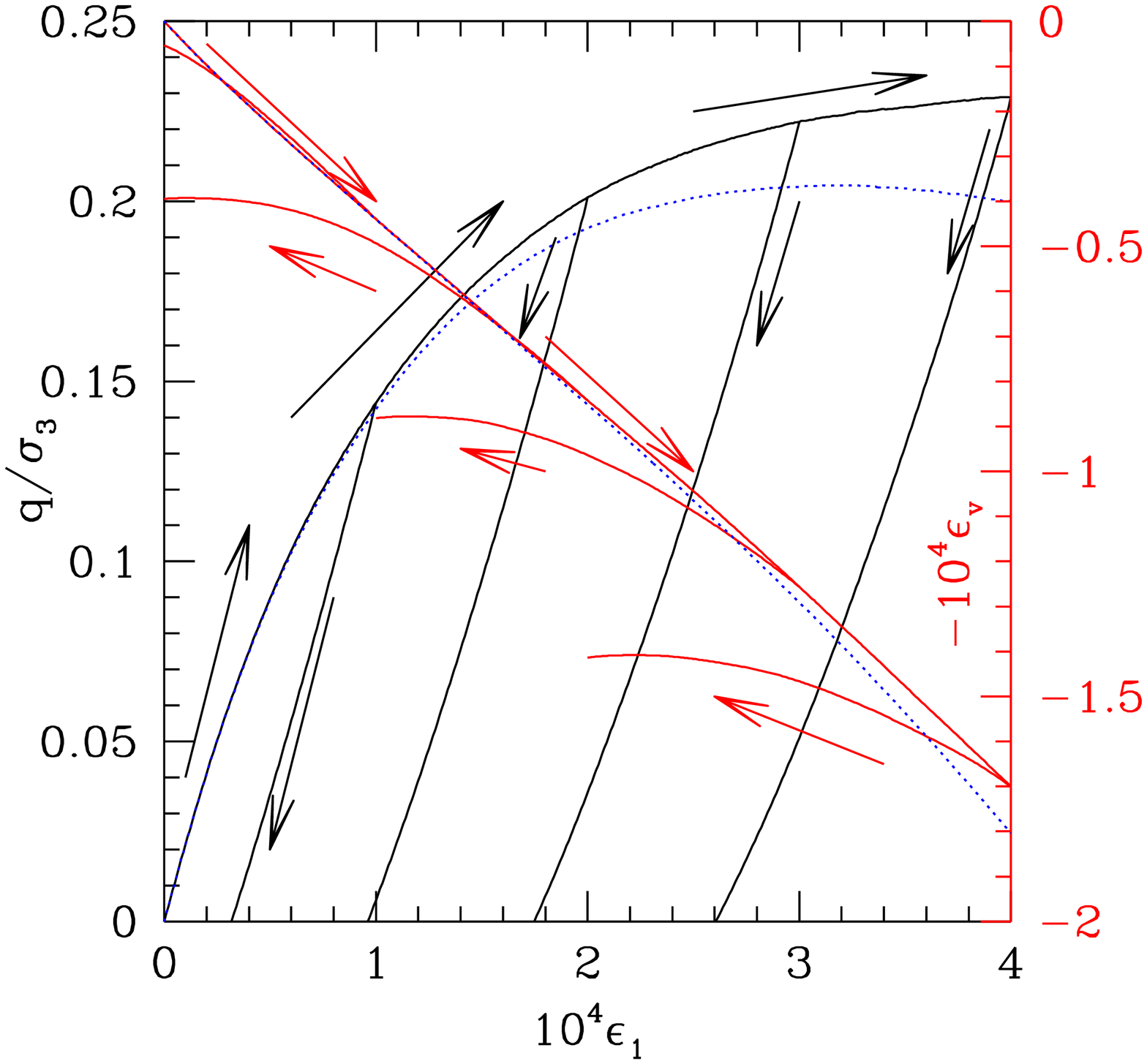}
\hfil
\includegraphics[width=.48\textwidth]{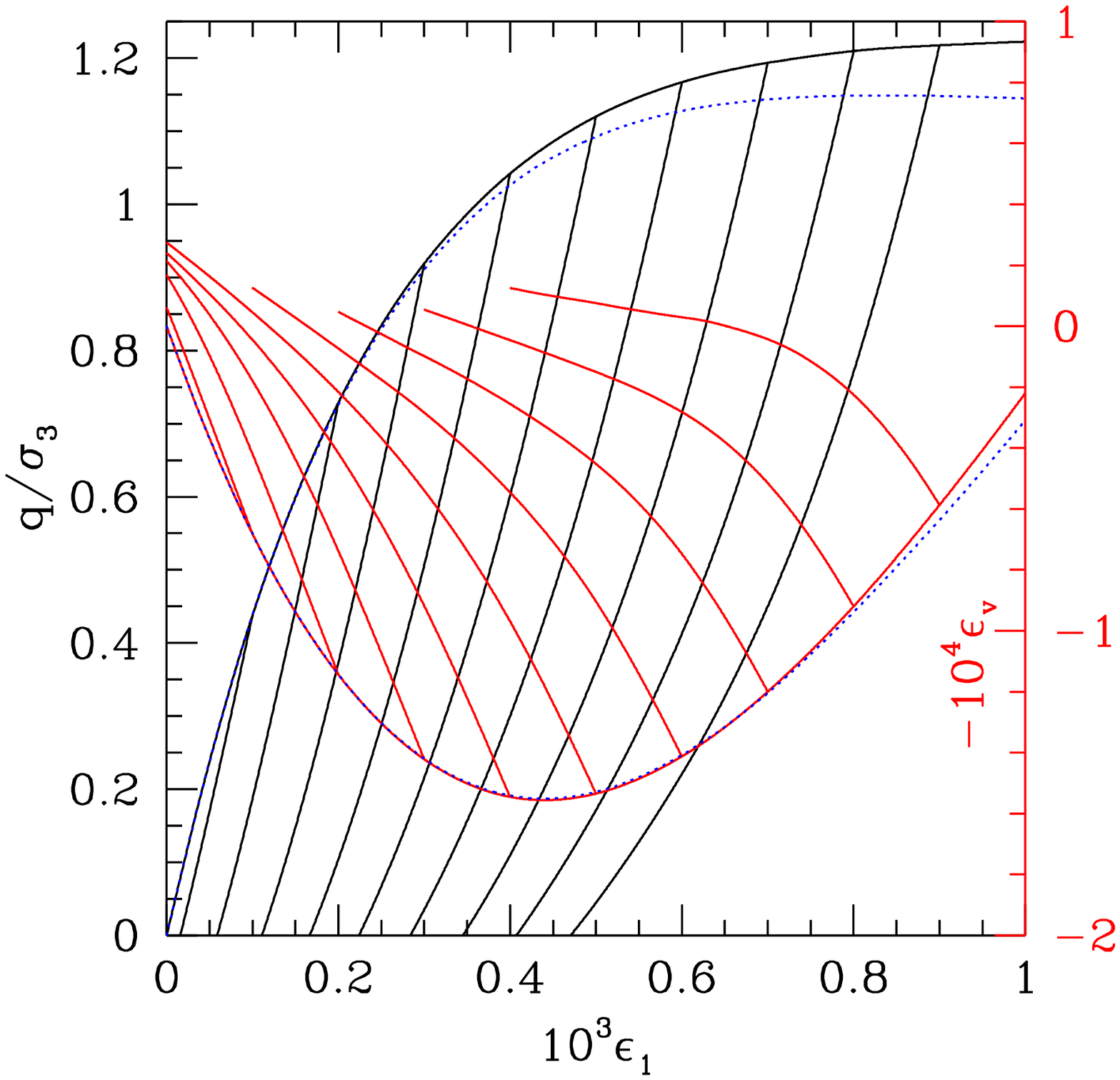}
\caption{\label{fig:cycle}
Courbes $q(\epsilon_1)/\sigma_3$ (axes de gauche) et  $\epsilon_v(\epsilon_1)$ (axes de droite), 
en compression triaxiale, simul\'ee par une m\'ethode de dynamique mol\'eculaire classique, 
pour le syst\`eme de la figure~\ref{fig:prq} \`a $P=100$~kPa (graphe de  gauche), et pour un syst\`eme
similaire mais avec une coordinence initiale beaucoup plus grande (graphe de droite). Noter les \'echelles de d\'eformation. 
Dans les deux cas, on a simul\'e les effets d'une d\'echarge \`a partir de diff\'erents \'etats atteints au cours de l'essai.
Les lignes en pointill\'es fins repr\'esentent les r\'esultats de la m\^eme simulation dans laquelle on a ignor\'e la cr\'eation
de nouveaux contacts.
}
\end{figure}

\begin{figure}[!htb]
\centering
\includegraphics[width=.48\textwidth]{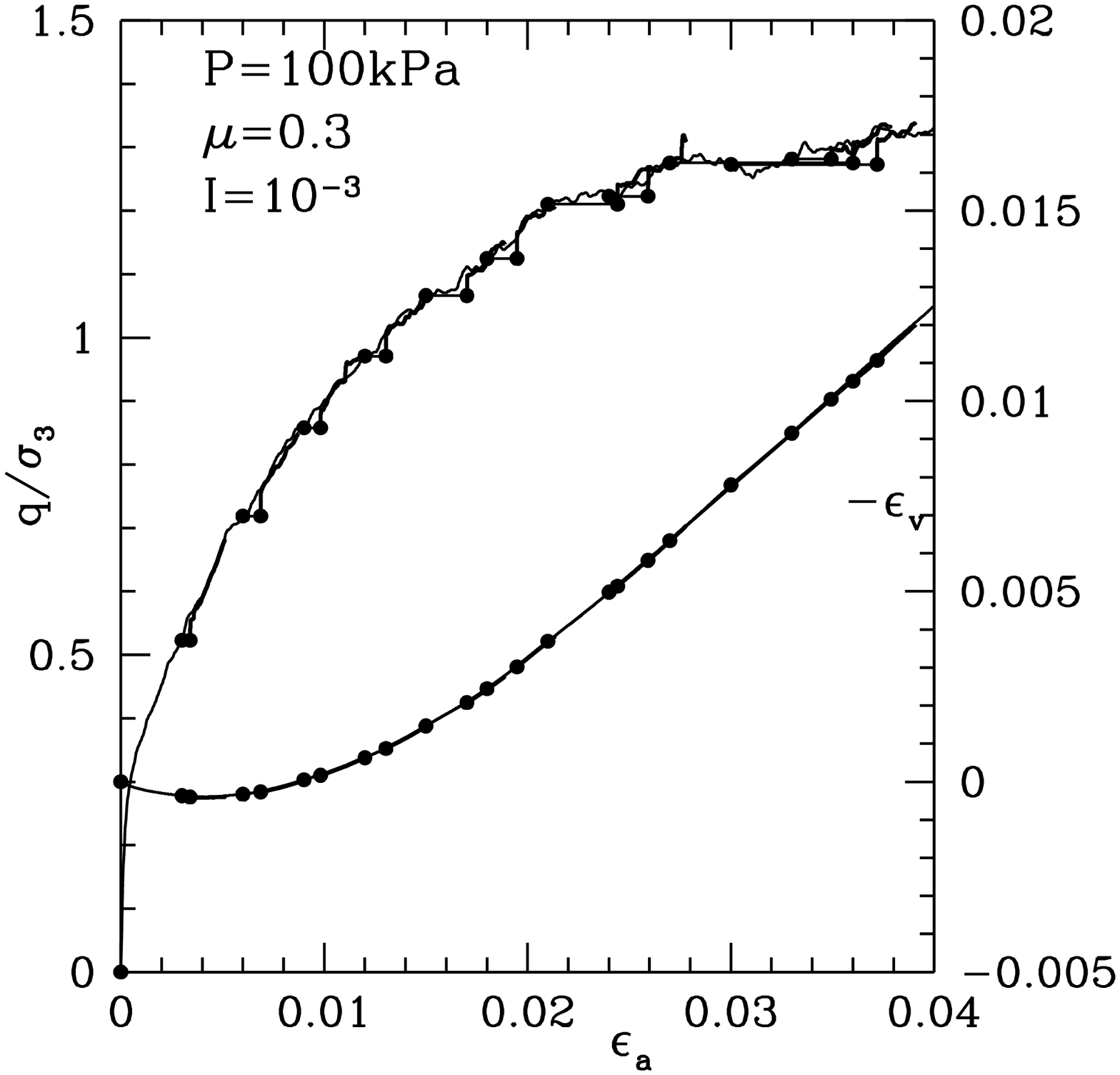}
\hfil
\includegraphics[width=.48\textwidth]{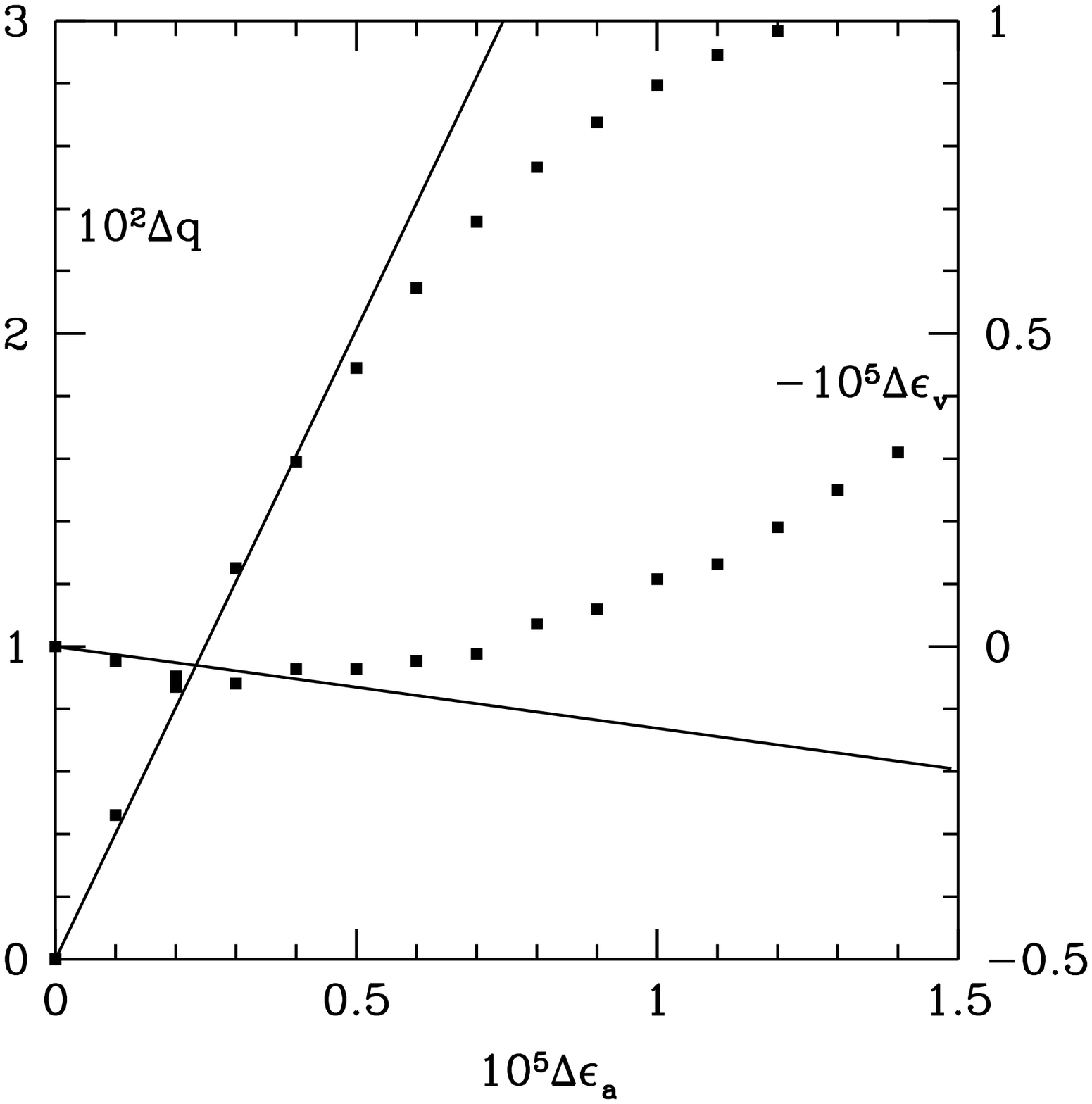}
\caption{\label{fig:dessJM3}
\'Equilibrage sous contraintes constantes, puis reprise de la compression triaxiale \`a $\dot\epsilon_1$ impos\'e \`a partir de diff\'erents \'etats interm\'ediaires.
Les symboles circulaires pleins montrent le d\'ebut et la fin de chacun des intervalles de fluage qui correspondent 
\`a ces \'equilibrages (graphe de gauche). La reprise
de l'essai \`a $\dot\epsilon_1$ impos\'e se caract\'erise par une r\'eponse initiale tr\`es raide (courbes en trait gras) dont le graphe de droite, 
analogue de la figure~\ref{fig:prq} pour un \'etat
d'\'equilibre apr\`es fluage le long de la trajectoire de l'essai triaxial, montre qu'elle est \'elastique.
}
\end{figure}

La figure~\ref{fig:prq} montre les variations du d\'eviateur des contraintes $q=\sigma_1-\sigma_3$, 
normalis\'e par la contrainte lat\'erale $\sigma_3$, et de la d\'eformation
volumique $\epsilon_v$, pr\`es de l'\'etat initial isotrope dans un essai de 
compression triaxiale de r\'evolution simul\'e pour des billes de verre, en fonction
de la d\'eformation axiale $\epsilon_a=\epsilon_1$. L'indice $1$ correspond ici \`a la direction principale majeure des contraintes.
Les r\'esultats sont montr\'es pour diff\'erentes valeurs de la contrainte isotrope initiale $P=\sigma_2=\sigma_3$. 
On voit que l'\'elasticit\'e au voisinage de l'\'etat initial d\'ecrit en 
bonne approximation la relation entre incr\'ements de d\'eformation et de contrainte pour des
d\'eformations de l'ordre de $10^{-5}$, et que ce domaine approximativement \'elastique augmente avec$P$. 
Hors du domaine de validit\'e de l'\'elasticit\'e lin\'eaire initiale, 
il est connu que la d\'eformation est irr\'eversible, comme le montre la figure~\ref{fig:cycle}.
Au cours de la compression triaxiale, on voit que la pente de la courbe de d\'eviateur d\'ecro\^{\i}t progressivement. Sous l'effet d'une d\'echarge,
on retrouve une pente plus \'elev\'ee, proche du module \'elastique initial. En fait, la tangente \`a la courbe de d\'echarge, 
on peut le v\'erifier, co\"{\i}ncide bien avec le module \'elastique correspondant, que l'on peut \'evaluer en faisant appel \`a la matrice de raideur pour l'\'etat 
consid\'er\'e. Ceci n'est possible que si la configuration est bien \'equilibr\'ee. On doit donc, avant de calculer les modules par l'approche statique, laisser 
s'\'equilibrer l'\'etat interm\'ediaire obtenu au cours de la simulation dynamique, en imposant des contraintes constantes 
plut\^ot qu'en poursuivant la d\'eformation \`a vitesse contr\^ol\'ee. 
Comme le montre la figure~\ref{fig:dessJM3}, cet \'equilibrage s'accompagne d'un l\'eger fluage.
Si on reprend par la suite la compression \`a taux de d\'eformation axial impos\'e, on observe une remont\'ee plus raide de la courbe de d\'eviateur, 
dont la pente initiale co\"{\i}ncide avec le module \'elastique calcul\'e par la matrice de raideur dans l'\'etat d'\'equilibre atteint apr\`es fluage.
Cette observation s'explique par la formation, lors de l'\'equilibrage, d'une structure plus stable avec 
les forces de contact strictement \`a l'int\'erieur du c\^one de Coulomb, 
sous l'effet des vibrations du syst\`eme autour de sa position finale, qui brouillent 
l'effet de polarisation des forces tangentielles du taux de d\'eformation maintenue
auparavant dans une direction constante. Une fois l'essai repris \`a $\dot\epsilon_1$ fix\'e, 
apr\`es une phase initial,les courbes de $q$ et $\epsilon_v$ rejoignent celles 
de l'essai monotone et sans arr\^et. Il est int\'eressant de noter que les modules 
\'elastiques se mesurent exp\'erimentalement de mani\`ere similaire~\cite{GBDC03}~:
on applique de petits cycles de contraintes autour de la valeur \`a laquelle l'essai a \'et\'e interrompu, 
ce qui provoque d'abord un certain fluage, puis les caract\'eristiques
\'elastiques se d\'eduisent de la forme finale stabilis\'ee et peu dissipative du cycle de contraintes et d\'eformations. 
En laboratoire c'est davantage la sollicitation appliqu\'ee qui est responsable du fluage
que l'\'ecart  \`a l'\'equilibre\footnote{On observe aussi un certain fluage dans les 
exp\'eriences mais seulement sur des dur\'ees beaucoup plus longues. On l'attribue au bruit ambiant. Voir le chapitre 9 pour une 
comparaison des valeurs de $\dot\epsilon$ entre les exp\'eriences et les simulations num\'eriques.}.

Pour conclure,
nous retiendrons de ces rappels rapides de r\'esultats et d'observations num\'eriques des 
propri\'et\'es \'elastiques des assemblages granulaires mod\`eles que leur
\'etude est facilit\'ee et syst\'ematis\'ee lorsque l'on a 
recours \`a l'approche quasi-statique fond\'ee sur la matrice de raideur. 
Les conditions dans lesquelles on observe une r\'eponse approximativement \'elastique et 
lin\'eaire sont tr\`es similaires dans la simulation et dans les exp\'eriences~: 
dans un intervalle du m\^eme ordre en tr\`es faible d\'eformation suite au processus
d'assemblage ; en d\'echarge lors d'un essai triaxial  : ou bien, sur de tr\`es 
courts intervalles, lors de la reprise de la d\'eformation monotone dans la m\^eme direction
suite \`a un arr\^et d'un essai \`a $\dot\epsilon$ impos\'e et \`a un petit fluage provoqu\'e par 
l'approche de l'\'equilibre ou par de faibles sollicitations cycliques. En dehors
du r\'egime approximativement \'elastique, on note que les d\'eformations restent de 
l'ordre de la pr\'ediction de l'\'elasticit\'e lin\'eaire d'autant plus longtemps que
le r\'eseau des contacts initiaux est bien connect\'e (figure~\ref{fig:cycle}).

\section{Applications \`a la d\'eformation an\'elastique\label{sec:plast}}
\subsection{Formulation du probl\`eme, propri\'et\'es g\'en\'erales}
Au-del\`a de la matrice de raideur \'elastique initiale, qui ne fournit que la tangente \`a 
l'origine des courbes rh\'eologiques, voyons comment et dans quels cas
l'approche quasi-statique peut pr\'edire la d\'eformation d'un \'echantillon granulaire 
soumis \`a un trajet de chargement donn\'e. Pour fixer les id\'ees nous traitons 
du cas simple de la compression biaxiale d'un \'echantillon bidimensionnel de disques, 
l'\'elasticit\'e du contact faisant intervenir des raideurs $K_N$ et  $K_T$ constantes.
Nous pr\'esentons et commentons ici les r\'esultats essentiels de nos propres travaux \cite{Gael2,RC02,CR03} 
sur ce probl\`eme particulier, sachant que des \'eclairages tr\`es utiles sont fournis
par les \'etudes de McNamara \emph{et al.}~\cite{McGaHe05,McHe06}. On traite le syst\`eme dans l'HPP, on n\'eglige $\stib$, 
pour des contrainte appliqu\'ees de la forme $\sigma_1 = P+q$, $\sigma_2 = P$, 
$P$ \'etant la pression initiale isotrope. Le probl\`eme consiste donc \`a d\'eterminer
le vecteur d\'eplacement $\dep$ et les forces de contact ${\bf f}$, tandis que le d\'eviateur $q$ augmente graduellement \`a partir de z\'ero.
Le vecteur $\fext$ ne contient que des valeurs nulles, sauf les coordonn\'ees qui correspondent aux contraintes, qu'elles s'expriment
par des forces sur les parois ($\sigma_1 = F_1/L_2$, $\sigma_2 = F_2/L_1$, voir la figure~\ref{fig:biax}) ou bien, dans le cas d'une cellule p\'eriodique, par
$(A\sigma_\alpha)_{\alpha = 1, 2}$ qui doit satisfaire~\eqref{eq:lovebiax}. 
En supposant que la trajectoire quasi-statique a \'et\'e trouv\'ee depuis l'\'etat de d\'epart 
jusqu\`a une certaine valeur de $q$, pour laquelle le vecteur d\'eplacement est $\dep(q)$, 
on cherche, pour un petit incr\'ement de chargement $\Delta\fext$ correspondant
\`a $\Delta q$ le surcro\^{\i}t de d\'eplacement $\Delta\dep$, solution de~:
\be
\sti (\dep(q),\Delta\dep)\cdot\Delta\dep = \Delta\fext.
\label{eq:problemD}
\ee
Dans~\eqref{eq:problemD}, la matrice de raideur $\sti$ d\'epend de ${\bf U}$ et aussi de la \emph{direction} de $\Delta\dep$ -- on pourrait argumenter $\sti$ par 
${\disty \frac{\Delta\dep}{\normm{\Delta\dep}}}$. On peut \'egalement \'ecrire~\eqref{eq:problemD} en faisant appara\^{\i}tre les d\'eriv\'ees par rapport
\`a $q$, qui joue le r\^ole d'un <<~temps cin\'ematique~>>, c'est-\`a-dire d'un param\`etre le long de la trajectoire dans l'espace des configurations~:
\be
\sti (\dep,\frac{d\Delta\dep}{dq})\cdot\frac{d\Delta\dep}{dq} = \frac{d\fext}{dq},
\label{eq:problemd}
\ee
Comme on a une \'elasticit\'e de contact unilat\'erale, la matrice de raideur est modifi\'ee avant tout par
la mobilisation du frottement, le bloc $\kk_{ij}$ prenant alors l'une ou l'autre des deux formes~\eqref{eq:raidplas}. 
Elle est aussi affect\'ee par l'ouverture des contacts -- lorsque
les grains $i$ et $j$ sont s\'epar\'es, il faut bien s\^ur prendre  $\kk_{ij}=0$. 
Au cours de l'\'evolution quasi-statique des d\'eplacements et des forces, un certain nombre
$\tilde N_c$ de contacts deviennent << critiques >> c'est-à-dire que le frottement y est compl\`etement mobilis\'e. 
\`A une \'etape donn\'ee du calcul, 
pour r\'esoudre \eqref{eq:problemD} il faut d\'eterminer le statut, glissant ou non glissant, de ces contacts critiques~: a priori la matrice de raideur peut donc
prendre $2^{\tilde N_c}$ formes diff\'erentes, et l'existence et l'unicit\'e de la solution posent probl\`eme. Une propri\'et\'e essentielle a \'et\'e \'etablie dans
\cite{McHe06}~: tant que l'ensemble des $2^{\tilde N_c}$ matrices de raideur satisfont le crit\`ere de stabilit\'e, 
c'est-à-dire la positivit\'e stricte de la forme quadratique
\eqref{eq:tso}, alors la solution existe et est unique. En pratique, plus la population de contacts glissants augmente, plus la stabilit\'e est menac\'ee. 
Lorsque certains 
statuts de contact pourraient donner $\Delta^2 W < 0$ avec certaines directions de $\delta\dep$, on peut s'attendre \`a la manifestation d'une instabilit\'e, un bruit
arbitrairement faible pouvant solliciter le syst\`eme dans la direction instable et d\'eclencher une augmentation exponentielle de l'\'energie cin\'etique. 
Il semble donc, gr\^ace au r\'esultat de~\cite{McHe06}, que l'approche quasi-statique fournisse une solution unique aussi longtemps qu'elle est physiquement fond\'ee.
Cette conclusion peut \^etre nuanc\'ee quelque peu cependant, car la d\'emonstration de~\cite{McHe06} ignore l'ouverture des contacts. \`A noter
que  ce ph\'enom\`ene n'est pas analogue au changement de statut~: en effet, alors qu'un contact glissant 
peut devenir non-glissant \`a tout instant si la force qu'il transmet
p\'en\`etre \`a l'int\'erieur du c\^one de frottement, un contact ouvert ne se referme que si un incr\'ement 
de d\'eplacement fini vient combler l'interstice qui est apparu.
L'ouverture des contacts est responsable (dans le cas lin\'eaire) de la d\'ependance en $\dep$ de $\sti$ dans~\eqref{eq:problemD},
alors que le changement de statut, de glissant \`a non glissant ou vice-versa, est la cause de sa d\'ependance dans la direction de l'incr\'ement $\Delta\dep$.

Tant que~\eqref{eq:problemd} peut d\'eterminer une trajectoire comme une suite d'\'etats d'\'equilibre, on a affaire \`a un r\'eseau d'\'el\'ements rh\'eologiques
(ressorts et patins frottants avec condition de Coulomb), la forme des grains n'intervient pas directement
(pourvu que l'approximation HPP soit bonne, ce qui est effectivement le cas sauf pour des grains anormalement mous), 
sauf \`a conditionner initialement la g\'eom\'etrie du r\'eseau de contacts. 
Il en r\'esulte que les d\'eformations, sous des contraintes donn\'ees, seront pour une m\^eme 
g\'eom\'etrie inversement proportionnelles aux raideurs $K_N$, $K_T$. 
Nous qualifions ce comportement de r\'egime I ou r\'egime strictement quasi-statique. Il arrive aussi
que l'\'evolution d'une collection de grains, qui reste proche de l'\'equilibre, 
se fasse par une succession de petits r\'earrangements~\cite{CR2000,RC02,CR03}, qui sont
d\'eclench\'es par des instabilit\'es. Celles-ci ne donnent toutefois que des mouvements de faible amplitude, 
arr\^et\'es par la fermeture de nouveaux contacts. Pour
des syst\`emes de taille croissante, l'intervalle de contrainte (de $q$ dans le cas de l'essai biaxial) 
est de plus en plus faible, ainsi que l'amplitude de
la d\'eformation qui accompagne le r\'earrangement du r\'eseau~\cite{RC02}, de sorte qu'\`a l'\'echelle 
macroscopique la d\'eformation appar\^{\i}tra comme graduelle
et continue. Nous qualifions ce comportement, dans la limite des \'evolutions macroscopiques lentes, 
o\`u l'inertie n'est plus pertinente\footnote{Voir au chapitre 9 
comment le r\^ole de l'inertie est \'evalu\'e par un nombre sans dimension.} 
de r\'egime quasi-statique au sens large ou r\'egime II -- les d\'eformations par
r\'earrangement \'etant dites << de type II~>>. Nous donnons ci-dessous (§\ref{sec:nonassoex}) 
quelques exemples d'utilisation de la m\'ethode quasi-statique, 
ainsi que quelques  illustrations des propri\'et\'es des r\'egimes I et II. Auparavant, 
voyons comment on m\`ene les calculs pour la r\'esolution de~\eqref{eq:problemd}.
\subsection{Un algorithme de calcul\label{sec:algo}}
\begin{figure}[!htb]
 \centering
%  \subfigure[Cas non associ\'e]{
  \psfrag{ft}{$T_{ij}$}
  \psfrag{fn}{$N_{ij}$}
  \psfrag{mu1}{$T_{ij} = \mu N_{ij}^N$}
  \psfrag{mu2}{$T_{ij} = -\mu N_{ij}^N$}
  \includegraphics[width=0.4\textwidth]{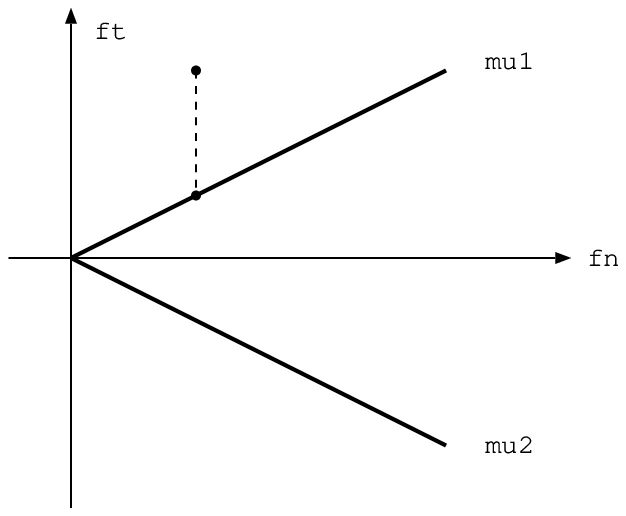}
%  \label{fig:projCoulomb}
%  }
  `\hfil
%  \subfigure[Cas associ\'e]{
  \psfrag{ft}{$\frac{T_{ij}}{\sqrt{K_T}}$}
  \psfrag{fn}{$\frac{N_{ij}}{\sqrt{K_N}}$}
  \psfrag{mu1}{\hskip -0.5cm pente $\mu \sqrt{\frac{K_N}{K_T}}$}
  \psfrag{mu2}{\hskip -1cm pente $-\mu \sqrt{\frac{K_N}{K_T}}$}
  \includegraphics[width=0.4\textwidth]{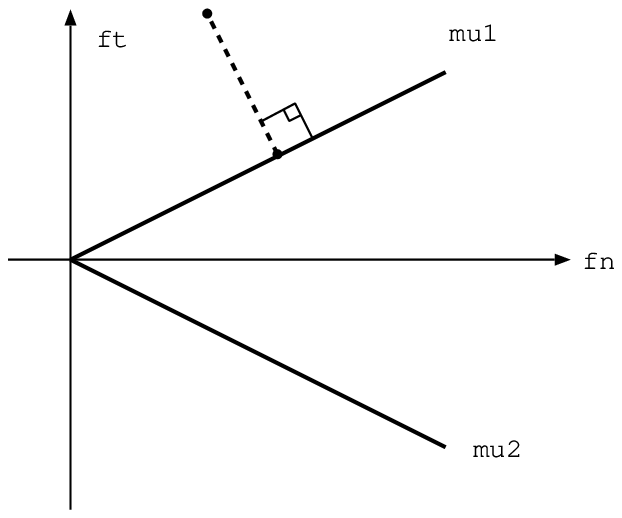}
%  \label{fig:projCoulombasso}
%  }
  \caption{\label{fig:projCou} Projection $\ppp$ sur le c\^one de Coulomb suivant
une r\`egle d'\'ecoulement non associ\'ee (graphe de gauche), ou  associ\'ee (graphe de droite).}
\end{figure}

Pour r\'esoudre num\'eriquement \eqref{eq:problemd}, 
on discr\'etise l'\'evolution du param\`etre de chargement $q$ en intervalle $\Delta q$ et on utilise la forme
incr\'ementale \eqref{eq:problemD}. 
Pla\c{c}ons-nous pour une valeur donn\'ee $q$, en supposant que le probl\`eme a \'et\'e r\'esolu pas \`a pas depuis $q=0$, ce qui
fournit les valeurs courantes de $\dep$ et de ${\bf f}$
et cherchons
\`a r\'esoudre it\'erativement \eqref{eq:problemD}.   
La m\'ethode de r\'esolution que nous r\'esumons ici se pr\'esente comme une recherche des incr\'ements de forces de contact $\Delta{\bf f}$, tels que 
${\bf f} + \delta{\bf f}$ soit plastiquement
et statiquement admissibles (voir le §~\ref{sec:defrole}), et de plus corresponde aux incr\'ements de d\'eplacement $\Delta\dep$, 
qui doivent satisfaire la condition suivante.
D\'efinissant les d\'eplacements relatifs \'elastiques par~:
$$
\Delta\deprel^{\text{E}} = (\kk^{\text{E}})^{-1}\cdot \Delta{\bf f}, 
$$
avec la forme \'elastique de la matrice des raideurs de contact,
et les d\'eplacements relatifs plastiques $\Delta\deprel^{\text{P}}$, par
$$
\Delta\deprel = \rigt\cdot\Delta\dep= \Delta\deprel^{\text{E}} + \Delta\deprel^{\text{P}},
$$
on doit avoir $\Delta\deprel^{\text{P}}_{ij}=0$ sauf pour les contacts glissants, 
où le d\'eplacement relatif plastique est (avec nos conventions) positif\footnote{Le 
vecteur unitaire tangentiel est tel que $T_{ij} >0$.}.
Cette condition s'exprime aussi simplement en notant que l'on a ${\bf f} + \kk\cdot\deprel = \ppp\left[{\bf f}+\kk^{\text{E}}\cdot\deprel \right]$ 
pour tout $·\deprel$, o\`u 
$\ppp$ d\'esigne le projecteur sur le c\^one de Coulomb, qui est d\'efini sur la figure~\ref{fig:projCou}. 
En plasticit\'e on dit qu'un tel projecteur d\'efinit la
\emph{r\`egle d'\'ecoulement}, c'est-\`a-dire le choix de la direction $\Delta\deprel^{\text{P}}$. Un autre choix serait de projeter orthogonalement
sur le c\^one, pour le produit scalaire qui corresponde \`a la norme $\norm{f}$ d\'efinie par
\be
\norm{f}^2 = {\bf f}\cdot(\kk^{\text{E}})^{-1}\cdot{\bf f}.
\label{eq:defnorm}
\ee
Ceci d\'efinit la r\`egle d'\'ecoulement dite \emph{associ\'ee} (figure~\ref{fig:projCou}, graphe de droite), alors que la r\`egle correspondant au
frottement de Coulomb usuel (graphe de gauche de la figure~\ref{fig:projCou}) est \emph{non associ\'ee}. Une loi d'\'ecoulement associ\'ee entra\^{\i}ne cette
propri\'et\'e remarquable que le calcul \`a la rupture fournit non seulement une condition n\'ecessaire, mais aussi suffisante de stabilit\'e
(\`a condition toutefois que l'on puisse n\'egliger l'influence de l'ouverture des contacts). Le r\'eseau de contact continuera de supporter le
chargement aussi longtemps qu'il existe des forces de contact qui soient \`a la fois statiquement et plastiquement admissibles. Pour le voir, il suffit
de v\'erifier que la solution $\Delta{\bf f}$ de~\eqref{eq:problemD} minimise $\norm{\Delta{\bf f}}^2$, avec la norme de~\eqref{eq:defnorm}, sous la contrainte
${\bf f}+\Delta{\bf f}\in \St$.

Pour r\'esoudre it\'erativement \eqref{eq:problemD}, on consid\`ere qu'on a affaire \`a un probl\`eme \'elastique corrig\'e par
l'application de forces ext\'erieures sur les grains concern\'es par les contacts glissants.
Au début du calcul, lorsque l'indice
d'itération $j$ vaut zéro, on prend 
$\Delta\dep_0=\sti^{-1}\cdot \Delta \fext$, c'est-à-dire la
solution élastique pour les incréments de déplacements.
On évalue alors les forces de contact comme~:
\be
{\bf f}_j = {\bf f}+\kk\cdot\rig\cdot\Delta\dep_j,
\label{eq:calfsa}
\ee
ce qui donne des forces statiquement admissibles.
Ces forces peuvent toutefois sortir du cône de Coulomb, aussi
leur applique-t-on la projection $\ppp$ pour obtenir des forces
plastiquement admissibles~:
\be
{\bf f}_j ^{PA}=\ppp \left[{\bf f}_j\right].
\label{eq:projfsa}
\ee
Le vecteur ${\bf f}_j ^{PA}$ n'est pas statiquement admissible, il
n'équilibre pas $\fext$ mais  
$$
\fext+\rigt \cdot\left({\bf f}_j ^{PA}-{\bf f}_j\right).
$$
On va par conséquent corriger le vecteur déplacement pour équilibrer
ces forces (en admettant le problème élastique). Cette correction vaut
\be
{\bf V}_j
=\sti^{-1}\cdot \rigt\cdot \left[ {\bf f}_j-{\bf
    f}_j^{PA}\right]
=\sti^{-1}\cdot \left[ \fext- \rigt\cdot{\bf f}_j ^{PA}\right] ,
\label{eq:calv}
\ee
et on incrémente $\Delta\dep$~:
\be
\Delta\dep_{j+1}=\Delta\dep_j+{\bf V}_j.
\label{eq:incru}
\ee
On peut alors remplacer $j$ par $j+1$, passant ainsi à l'itération
suivante qui reprend au calcul des forces par~\eqref{eq:calfsa}, et on
poursuit jusqu'à ce que ${\bf V}_j$ ou bien ${\bf f}_j ^{PA}-{\bf
  f}_j$ soit négligeable.

Cet algorithme, du point de vue des forces, revient à projeter
alternativement sur le cône de Coulomb avec $\ppp$, et sur l'espace
affine des forces statiquement admissibles avec $\qqq$ qui est une
projection orthogonale au sens de la norme~\eqref{eq:defnorm}. Les
opérations~\eqref{eq:calv} et \eqref{eq:incru} à l'étape $j$,
suivies de~\eqref{eq:calfsa} à l'étape $j+1$ se traduisent par~:
$$
{\bf f}_{j+1} = \qqq \left[ {\bf f}_j^{PA}\right].
$$

[Vérifions-le en définissant $ \qqq \left[ {\bf f}\right]$ pour un
${\bf f}$ quelconque. L'espace sur lequel il s'agit de projeter est
l'ensemble des ${\bf g}$ tels que $\rigt\cdot{\bf g}=\fext$. Il s'agit de décomposer ${\bf f}$ en $\qqq \left[ {\bf
    f}\right] + {\bf f}^{\prime}$, avec ${\bf f}^{\prime}$ orthogonal
au sens du produit scalaire associé à~\eqref{eq:defnorm} au directeur
de cet espace affine. En d'autres termes $\kk^{-1}\cdot {\bf
  f}^{\prime}$ doit être orthogonal au sens du produit scalaire
ordinaire entre déplacements relatifs et forces de contact, à tout
${\bf g}$ tel que $\rigt\cdot{\bf g}=0$, soit au noyau de $\rigt$.
L'orthogonal du noyau de $\rigt$ n'est autre que l'image de  $\rig$, on a 
$\kk^{-1}\cdot {\bf f}^{\prime}=-\rig \cdot {\bf V}$, où ${\bf V}$ est un certain
vecteur de déplacements. Pour le déterminer, il suffit d'écrire que
$\qqq  \left[{\bf f} \right]$ est statiquement admissible. Appliquant l'opérateur
$\rig$ aux deux membres de l'égalité 
\be
{\bf f} = \qqq  \left[{\bf f}
\right]-\kk\cdot\rig\cdot {\bf V},
\label{eq:decomp}
\ee
on trouve 
$${\bf V}=\sti^{-1}\cdot \left(\fext-\rigt\cdot{\bf
  f}\right),$$
et on en déduit $\qqq  \left[{\bf f} \right]$ par~\eqref{eq:decomp}.
Ce sont là exactement les opérations qui conduisent de ${\bf f}^{PA}_j$ à
${\bf f}_{j+1}$.]

Le m\^eme algorithme, dans le cas associé, pour lequel
$\ppp$ est un projecteur orthogonal au sens de la même
norme~\eqref{eq:defnorm} que $\qqq$, permet de retrouver la
propriété que si le
chargement est supportable (c'est-à-dire si $\St\ne\emptyset$),
alors on trouvera effectivement une solution au problème
élastoplastique, quelle que soit l'histoire du chargement. En effet il
s'agit alors de projeter orthogonalement alternativement sur deux
parties convexes fermées d'un espace de dimension finie, et la suite
obtenue doit converger vers un élément de leur intersection $\St$ si elle
n'est pas vide. 

\begin{figure}[!htb]
\centering
\includegraphics[angle=270,width=0.65\textwidth]{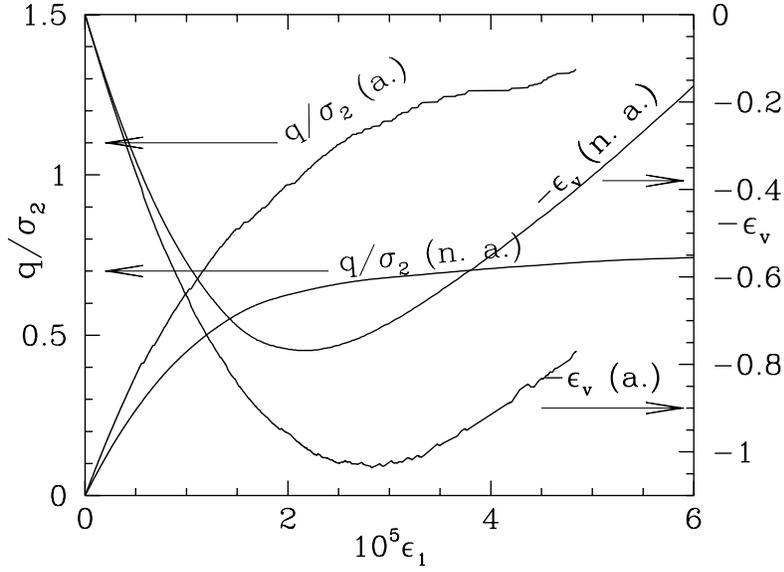}
\caption{\label{fig:compasso}
Comparaison des r\'egimes de chargement support\'e en d\'eformation strictement quasi-statique, pour la loi de contact habituelle, 
non associ\'ee (courbes marqu\'ees <<~n. a.~>>)et pour une loi de glissement associ\'ee (courbes marqu\'ees <<~a.~>>).
}
\end{figure}

Il n'en est pas de même avec la loi de glissement non-associ\'ee, car le d\'eviateur maximal que l'on atteint est
strictement inf\'erieur au r\'esultat du calcul associ\'e, comme le montre la figure~\ref{fig:compasso}~: le d\'eviateur maximal atteint
en r\'egime I  est voisin de $0,8$P dans le calcul non associ\'e et superieur \`a $1,3$P dans le calcul associ\'e.
Dans la mise en {\oe}uvre de la m\'ethode quasi-statique, on doit prendre en compte l'ouverture des contacts. En fait on peut le faire
dans le cadre it\'eratif de l'algorithme d\'ecrit ci-dessus, en gardant la matrice de raideur initiale et en faisant intervenir des forces auxiliaires
pour corriger l'effet de l'annulation de certains blocs $\kk_{ij}$. 
\subsection{Illustration~: r\'egimes I et II en compression biaxiale\label{sec:nonassoex}}
La m\'ethode quasi-statique permet de calculer l'\'evolution du syst\`eme sous le trajet de chargement biaxial jusqu'`a une
certaine valeur $q_1$ du d\'eviateur, qui borne l'intervalle de d\'eformation de type I et vaut environ $0,815P$  
dans l\'etude de~\cite{Gael2}, qui porte sur la compression triaxiale de syst\`emes de disques polydispers\'es assembl\'es initialement  
dans une configuration tr\`es dense (en fixant $\mu=0$ pendant l'assemblage, cf. le chapitre 8 de cet ouvrage), avec une grande raideur de contact ($K_N/P=10^5$)
et un coefficient de frottement $\mu=0,25$. 
$q_1$ ne diminue pas lorsque $N$ augmente (\`a la diff\'erence de 
l'intervalle de stabilit\'e des r\'eseaux de contact entre grains rigides non frottants~\cite{CR2000,RC02}). Il semble, au contraire 
augmenter l\'eg\`erement (les r\'esultats sont compatibles avec un effet de taille finie en $-(2,12)/\sqrt{N}$). Il est montr\'e par ailleurs dans~\cite{Gael2}
que $q_1$ est ind\'ependant de la raideur (si elle est assez grande, d'ordre $10^4$) et du rapport $K_T/K_N$. On observe par ailleurs qu'une proportion finie
de contacts adopte le statut glissant (jusqu'\`a 20\% dans~\cite{Gael2}) et que 5 \`a 10\% des contacts sont perdus. 
\`A noter que dans l'approche statique, aucune vibration <<~parasite~>> ne vient
brouiller la distinction entre contacts glissants et non glissants.

\begin{figure}[!htb]
\centering
\includegraphics[angle=270,width=0.75\textwidth]{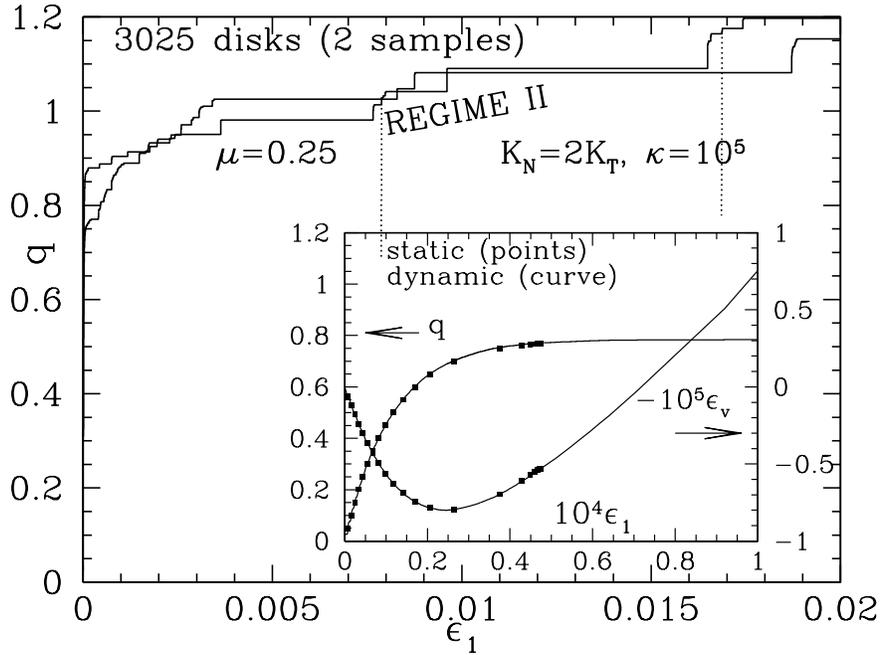}
\caption{\label{fig:dessmarfro}
D\'eviateur $q$ en fonction de la d\'eformation << axiale >> dans un essai biaxial 2D simul\'e \`a contraintes contr\^ol\'ees par
petits paliers.
En insert (noter la dilatation des \'echelles de d\'eformation) on montre l'intervalle $q\le q_1$ en r\'egime I, et on compare les calculs 
avec la m\'ethode quasi-statique avec les r\'esultats de simulation dynamique. 
}
\end{figure}

$q_1$ reste nettement inf\'erieure au maximum de d\'eviateur, $q_{\text{max}}$, d'o\`u un r\'egime de d\'eformation par r\'earrangements (type II)
pour $q_1\le q\le q_{\text{max}}$. La figure~\ref{fig:dessmarfro} illustre ces deux phases du comportement dans une compression biaxiale monotone.
Elle repr\'esente le r\'esultat de calculs par dynamique mol\'eculaire \`a contrainte contr\^ol\'ee, 
en imposant des pas de d\'eviateur $\Delta q = 10^{-3}P$, 
puis en attendant l'\'equilibre pour chaque valeur de $q$ avant de l'incr\'ementer \`a nouveau. Lorsque le r\'eseau initial des contacts reste stable,
on a des d\'eformations de type I d'ordre $K_N^{-1}$, comme les d\'eformations \'elastiques, et  
si faibles que la courbe se confond avec l'axe des ordonn\'ees sur le graphe. En dilatant l'\'echelle des d\'eformations (en insert sur la figure), 
on voit que dans ce r\'egime les calculs en dynamique mol\'eculaire et par la m\'ethode quasi-statique sont en excellent accord. Au-del\`a de $q_1$, on observe
une courbe $q(\epsilon_1)$ en forme d'escalier. Dans les phases de stabilit\'e
(parties d'allure verticale) la d\'eformation est \`a nouveau de type I et d'ordre $K_N^{-1}$. On a pu v\'erifier qu'un calcul par la m\'ethode quasi-statique
\'etait possible.  Dans les phases de r\'earrangement
(parties horizontales), le syst\`eme se d\'eforme par rupture du r\'eseau des contacts jusqu'\`a ce qu'un nouveau r\'eseau apparaisse et soit capable
de supporter le d\'eviateur appliqu\'e. \`A la diff\'erence des d\'eformations de type I, l'amplitude de
ces \'ev\'enements de rupture n'est pas li\'ee \`a la raideur des contacts. Les d\'eformations (de type II) qui en r\'esultent sont analogues aux d\'eformations
que l'on observe avec des mod\`eles de grains rigides, comme en dynamique des contacts (voir le chapitre 3 de ce trait\'e). La sensibilit\'e au niveau de raideur est d'ailleurs un 
moyen de d\'etecter le type de d\'eformation -- voir \`a ce propos la discussion de l'influence du niveau de raideur sur le comportement quasi-statique au chapitre 9.
Un autre moyen d'identifier, au moins approximativement, la nature (I ou II) des d\'eformations est de tester jusqu'o\`u il est possible, dans un calcul dynamique, 
de simuler, par exemple, un test biaxial ou triaxial, lorsque l'on ne cr\'ee aucun contact nouveau~: on \'eprouve alors la stabilit\'e du r\'eseau initial. Le
r\'esultat de tels calculs est montr\'e sur la figure~\ref{fig:cycle}~: l'intervalle de d\'eviateur en r\'egime I s'\'etend environ jusqu'\`a $q=0,2\sigma_3$
pour le syst\`eme initialement le moins coordonn\'e, et jusqu'\`a $q\simeq  1,1\sigma_3$ pour l'assemblage de coordinence plus \'elev\'ee. 
Il est naturel qu'un r\'eseau de contacts mieux connect\'e soit capable de supporter un intervalle de contraintes macroscopiques plus \'etendu.

\`A ce jour nous ne disposons pas d'analyse pr\'ecise des m\'ecanismes de rupture des r\'eseaux de contact pour $q=q_1$. C'est une perspective prometteuse
dans l'\'etude fine des m\'ecanismes de d\'eformation des assemblages granulaires (\`a rapprocher d'autres mat\'eriaux amorphes). 
\section{Conclusion\label{sec:conc}}
Quoique loin de concurrencer les m\'ethodes dynamiques, polyvalentes et d'emploi plus facile, les approches quasi-statiques, fond\'ees sur la construction
de matrices de raideur, sont de pr\'ecieux outils d'analyse des assemblages granulaires solides, \`a l'equilibre ou en d\'eformation quasi-statique. D'un point de vue
pratique, la construction de ces matrices fournit des moyens commodes pour juger de la stabilit\'e des configurations d'\'equilibre et pour \'evaluer leurs
propri\'et\'es \'elastiques. L'\'etude des matrices de rigidit\'e et de raideur met en lumi\`ere les influences des diff\'erentes donn\'ees 
g\'eom\'etriques et m\'ecaniques et fournit d'utiles indications sur le nombre de coordination. 
Hors du petit domaine de r\'eponse approximativement \'elastique, l'approche quasi-statique montre l'existence de deux
r\'egimes de comportement  caract\'eris\'es par des origines physiques distinctes de la d\'eformation macroscopique 
et des sensibilit\'es diff\'erentes aux param\`etres microm\'ecaniques. 
Encore assez embryonnaire, l'usage de l'approche quasi-statique et des matrices de raideur devrait trouver des applications fructueuses dans 
les \'etudes pr\'ecises des m\'ecanismes de d\'eformation des assemblages granulaires par instabilit\'e, rupture et r\'earrangement \`a 
l'\'echelle microscopique. Comment se comporte la distinction entre r\'egimes I et II dans la limite des grands syst\`emes et dans la limite des grains rigides~?
Avec des grains frottants de forme non sph\'erique, existe-t-il des m\'ecanismes stables, sources de <<~modes mous~>> dans le spectre de vibration~?
Quelle est l'allure \`a grande \'echelle des champs de d\'eplacements lors du d\'eclenchement de la rupture~? 
Comment le processus de d\'eformation par rupture d\'epend-il de la forme des grains~? 
Telles sont certaines des questions assez fondamentales
que le d\'eveloppement des m\'ethodes quasi-statiques devrait permettre de clarifier.

%-----------------------------------------------------------------------------------------------------------

\bibliographystyle{unsrt}

\bibliography{chapter1_biblio}

%-----------------------------------------------------------------------------------------------------------

\end{document}